\documentclass[15pt,a4paper]{article}
\usepackage[square, numbers]{natbib} 
\usepackage[english]{babel}
\usepackage[utf8]{inputenc}
\usepackage[T1]{fontenc}
\usepackage{amsmath,amssymb,amsfonts,amsthm}
\usepackage[fleqn]{nccmath}
\usepackage[tbtags]{mathtools}
\usepackage{dsfont} 
\usepackage{authblk}
\usepackage{graphicx}
\usepackage{xcolor}
\usepackage{float}
\usepackage{geometry}
\usepackage{booktabs}
\usepackage{multirow}
\usepackage{xr-hyper} 
\usepackage[colorlinks=true, urlcolor=blue, linkcolor=red]{hyperref}
\usepackage{array}
\usepackage[labelfont=bf]{caption}
\usepackage{subcaption}
\usepackage{floatpag} 
\floatpagestyle{empty} 
\usepackage{setspace} 
\usepackage{mathpazo} 

\usepackage[normalem]{ulem} 

\usepackage[ruled,vlined]{algorithm2e}

\DeclareMathOperator{\Var}{\mathbb{V}\text{ar}}

\DeclareMathOperator{\Tr}{Tr}

\DeclareMathOperator{\E}{\mathbb{E}}

\DeclareMathOperator{\sign}{sign}
\DeclareMathOperator{\diag}{diag}

\newcommand{\argmin}{\operatornamewithlimits{argmin}}
\newcommand{\transpose}{{}^{\text{\sffamily T}}}
\newcommand{\bigO}{O}
\newcommand{\smallO}{o}

\newtheorem{theorem}{Theorem}
\newtheorem{lemma}{Lemma}
\newtheorem{proposition}{Proposition}
\newtheorem*{remark}{Remark}

\makeatletter
\newcommand*{\addFileDependency}[1]{
  \typeout{(#1)}
  \@addtofilelist{#1}
  \IfFileExists{#1}{}{\typeout{No file #1.}}
}
\makeatother

\title{AdapDISCOM: An Adaptive Sparse Regression Method for High-Dimensional Multimodal Data With Block-Wise Missingness and Measurement Errors}

\author[1,2]{ Maimouna Baldé}
\author[1,2]{Abdoul O. Diakité}
\author[2,3]{Claudia Moreau}
\author[8]{Gleb Bezgin}
\author[4]{Nikhil Bhagwat}
\author[5,6,7]{Pedro Rosa-Neto}
\author[4]{Jean-Baptiste Poline}
\author[2,3]{Simon Girard}
\author[1,2]{Amadou Barry\footnote{Corresponding author: amadoudiogo.barry@inrsc.ca.} }
\author[**]{for the Alzheimer’s Disease Neuroimaging Initiative}

\affil[1]{Institut national de la recherche scientifique (INRS), Centre Armand-Frappier Santé Biotechnologie, Laval, Canada}
\affil[2]{Unité mixte de recherche (UMR) INRS-UQAC en santé durable, Chicoutimi, Canada}
\affil[3]{Université du Québec à Chicoutimi (UQAC), Chicoutimi, Canada}
\affil[4]{NeuroDataScience-ORIGAMI lab, McConnell Brain Imaging Centre, The Neuro (Montreal Neurological Institute-Hospital), Faculty of Medicine and Health
Sciences, McGill University, Montreal, Quebec, Canada}

\affil[5]{Translational Neuroimaging Laboratory, McGill University Research Centre for Studies in Aging, McConnell Brain Imaging Centre (BIC), Montreal Neurological Institute, Montreal Neurological Institute-Hospital, Montreal, QC, Canada}
\affil[6]{Institution: Douglas Hospital Research Centre - Centre intégré universitaire de santé et services sociaux de l’Ouest-de-l’Île-de-Montréal, Verdun, Quebec, Canada} 
\affil[7]{ The Peter O’Donnell Jr. Brain Institute (OBI), University of Texas Southwestern Medical Centre (UTSW). Dallas, TX, USA}
\affil[8]{Department of Neurology and Neurosurgery, Faculty of Medicine and Health Sciences, McGill University, Montreal, Quebec, Canada} 

\affil[**]{Data used in preparation of this article were obtained from the Alzheimer’s Disease Neuroimaging Initiative
(ADNI) database (adni.loni.usc.edu). As such, the investigators within the ADNI contributed to the design
and implementation of ADNI and/or provided data but did not participate in analysis or writing of this report.
A complete listing of ADNI investigators can be found at:
\url{http://adni.loni.usc.edu/wp-content/uploads/how_to_apply/ADNI_Acknowledgement_List.pdf } } 

\date{\today}

\begin{document}
\maketitle

\section{Abstract}
Multimodal high-dimensional data are increasingly prevalent in biomedical research, yet they are often compromised by block-wise missingness and measurement errors, posing significant challenges for statistical inference and prediction. We propose AdapDISCOM, a novel adaptive direct sparse regression method that simultaneously addresses these two pervasive issues. Building on the DISCOM framework, AdapDISCOM introduces modality-specific weighting schemes to account for heterogeneity in data structures and error magnitudes across modalities. We establish the theoretical properties of AdapDISCOM, including model selection consistency and convergence rates under sub-Gaussian and heavy-tailed settings, and develop robust and computationally efficient variants (AdapDISCOM-Huber and Fast-AdapDISCOM). Extensive simulations demonstrate that AdapDISCOM consistently outperforms existing methods such as DISCOM, SCOM, and CoCoLasso, particularly under heterogeneous contamination and heavy-tailed distributions. Finally, we apply AdapDISCOM to Alzheimer’s Disease Neuroimaging Initiative (ADNI) data, demonstrating improved prediction of cognitive scores and reliable selection of established biomarkers, even with substantial missingness and measurement errors. AdapDISCOM provides a flexible, robust, and scalable framework for high-dimensional multimodal data analysis under realistic data imperfections.
\clearpage

\section{Introduction}
Multimodal, high-dimensional data are increasingly pervasive in modern scientific research, underpinning advances in fields such as health, genetics, and neuroscience. By integrating complementary information from diverse modalities, such datasets enable richer modeling and more accurate prediction, supporting personalized medicine, large-scale genetic studies, and drug discovery \citep{Garca2014BiobanksAT}. However, their analysis presents significant statistical and computational challenges, particularly when data are subject to block-wise missingness and measurement errors \textemdash two common imperfections in real-world settings.

In biobanks like the Alzheimer’s Disease Neuroimaging Initiative (ADNI) study \citep{Mueller2005TheAD}, which integrates clinical, genetic, and neuroimaging data (structural, functional, molecular), block-wise missingness arises when entire modalities are unavailable for some individuals\textemdash often due to high costs (e.g., PET scans), technical failures, or patient reluctance (e.g., CSF collection). Simultaneously, the observed data are frequently contaminated with additive measurement errors, whose magnitude and nature vary across modalities depending on acquisition techniques. Ignoring these issues risks producing biased estimates, poor predictive performance, and unreliable variable selection. Standard statistical methods and algorithms fail to address these challenges effectively. Therefore, developing innovative methods and algorithms to mitigate the influence of measurement errors and missing data is crucial, especially given the size and complexity of modern datasets.

In recent years, substantial research efforts have been devoted to reducing the impact of measurement error and missing values in high-dimensional data. Regarding measurement errors, \citet{loh_Wainwright_2012} introduced a nonconvex modification of the Lasso objective function \citep{lasso_tib}, incorporating a constrained parameter that adds an optimization constraint. Despite providing theoretical guarantees of convergence close to the global minimum, the nonconvex nature of the objective function complicates computations. To address this limitation, \citet{datta_cocolasso_2017} proposed the Convex Conditioned Lasso (CoCoLasso), utilizing a positive semi-definite covariance estimator to achieve similar theoretical and computational advantages as the Lasso. Subsequent improvements have been proposed, such as BD-CoCoLasso \citep{escribe2021block}, which reduces the computational burden, and balanced estimation techniques \citep{ZHENG_balanced_2018}, which introduce an $l_0$ penalty to CoCoLasso for improved variable selection and prediction trade-offs. More recently, \citet{tao2021calibrated} proposed the CaZnRLS estimator, which calibrates least squares loss with a positively defined projection of an unbiased substitute for the covariate covariance matrix. A comprehensive review of recent advances in measurement error correction methods for high-dimensional data can be found in \citep{luo2023overview}.

Notably, CoCoLasso and its extensions primarily provide separate solutions for multiplicative measurement errors, where missing data are treated as a particular case. Meanwhile, recent methodologies have focused explicitly on handling missing data in high-dimensional settings. Many of these methods depart from traditional imputation approaches, which generate unstable matrices, accumulate imputation errors, and are computationally expensive for large-scale data. For instance, \citet{Scom2023} proposed a sample-wise combined missing effect model (SCOM) with penalization, integrating missing data from each sample into the model specification as an estimable parameter. \citet{yu_optimal_2020} developed a direct sparse regression procedure using covariance  from multimodality (DISCOM) method, inspired by \citet{ledoit_well-conditioned_2004}'s covariance estimator, to account for missing data in multimodal high-dimensional datasets. DISCOM estimates covariance as a linear combination of the identity matrix, intra-modal covariance estimates, and cross-modal covariance estimates. This covariance estimator is then incorporated into Lasso optimization for sparse estimation of optimal linear prediction coefficients. A key advantage of DISCOM is its ability to leverage all available information, as it only requires the minimum number of samples with at least two observed modalities, significantly increasing the effective sample size compared to complete-case analysis. DISCOM has gained considerable interest, with recent adaptations for multi-omics network data \citep{henao_multi-omics_2023}, accelerated failure time models with censored responses \citep{Wang2022RegularizedBM}, and multi-response variable settings \citep{Wang2024MultiresponseRF}. Additionally, \citet{He2023VariableSF} integrated DISCOM into an imputation algorithm for generalized linear models, further extending its applicability.

While the literature is rich in methods addressing measurement errors and missing data, these approaches typically handle each issue separately. However, in practice, both problems often coexist, particularly in large-scale, multimodal, high-dimensional datasets. To the best of our knowledge, no existing methods explicitly address the simultaneous presence of measurement errors and missing data.

In this manuscript, we present AdapDISCOM—an adaptive direct sparse regression procedure leveraging multimodal covariance—a novel method specifically designed to jointly address measurement errors and missing data in high-dimensional multimodal datasets. Building on the DISCOM framework under the missing completely at random assumption, AdapDISCOM introduces adaptive mechanisms through modality-specific weighting of the covariance structure, thereby accounting for heterogeneity in both data-generating processes and noise magnitudes across modalities. Such noise can vary substantially between data sources, as sequencing technologies, medical imaging scanners, and clinical assessments each introduce distinct types and magnitudes of error. Unlike traditional imputation-based or bias-correction methods, AdapDISCOM operates directly on the observed incomplete data, exploiting all available information and maintaining scalability. We also propose a computationally efficient strategy for tuning these weights.

The key contributions of this manuscript are as follows: (i) demonstrating that DISCOM mitigates the impact of additive measurement errors; (ii) proving that AdapDISCOM effectively handles both measurement errors and missing data simultaneously, addressing a more realistic data scenario; (iii) showing that AdapDISCOM accounts for the heterogeneity and varying intensity of measurement errors across different modalities; (iv) establishing the theoretical properties of AdapDISCOM estimators, including model selection consistency; and (v) empirically validating the superiority of AdapDISCOM through realistic simulations, where data generation and measurement errors differ across modalities.

The remainder of this manuscript is organized as follows. We first present the DISCOM and AdapDISCOM methods, demonstrating their ability to jointly address measurement errors and missing data in high-dimensional multimodal datasets. Next, we establish the theoretical properties of AdapDISCOM estimators, including model selection consistency, and introduce an efficient hyperparameter selection approach. We then provide extensive simulation studies, incorporating realistic scenarios where each modality follows distinct data generation and noise processes. Finally, we illustrate the application of AdapDISCOM to real-world datasets before concluding with a discussion of future research directions.

\section{Method}\label{sec_method}
We first describe the DISCOM approach for multimodal data and demonstrate its ability to mitigate the effects of missing values and additive measurement errors. We then introduce our proposed method, AdapDISCOM, which generalizes DISCOM and provide its theoretical properties. Additionally, we propose two extensions: a robust version, referred to as AdapDISCOM-Huber, based on the Huber's M-estimate for the heavy-tailed case, and a computationally efficient variant, Fast-AdapDISCOM, which reduces the number of hyperparameters to be estimated.

In the following, vectors are written in lower bold letters $(\boldsymbol{x}\in\mathbb{R}^p),$ and matrices in capital bold letters $(\boldsymbol{X}\in\mathbb{R}^{n\times p}).$ Estimated quantities are represented with a symbol above the characters like $(\widehat{\boldsymbol{X}}, \ \widetilde{\boldsymbol{X}}, \ \check{\boldsymbol{X}}),$ the inverse of a squared matrix is noted as $\boldsymbol{X}^{-1}$ and the transposed matrix is $\boldsymbol{X}\transpose.$ For a matrix we use $ \| \boldsymbol{X} \|_{F}, \ \|\boldsymbol{X}\|_{\max},$ and  $\|\boldsymbol{X} \|_{\infty}$ to denote the Frobenius norm $\sqrt{\sum_{ij}x_{ij}^2},$ the max norm $\max_{ij}|x_{ij}|,$ and the infinity norm $\max_{i}\sum_{j}^p|x_{ij}|,$ respectively. For a vector we use $\| \boldsymbol{x} \|_{2}, \ \|\boldsymbol{x}\|_{\max},$ and  $\|\boldsymbol{x} \|_{1}$ to denote the $l_2$ norm $\sqrt{\sum_{j}x_{j}^2},$ the max norm $\max_{j}|x_{j}|,$ and the $l_1$ norm $\sum_{j}^p|x_{j}|,$ respectively.  

\subsection{DISCOM accounting for missing data and additive measurement error}\label{desc_metho}
We want to predict the centered response variable $\textbf{y}$ from the linear regression model $\textbf{y} = \textbf{X}\boldsymbol{\beta} + \boldsymbol{\varepsilon},$ where $\textbf{X} = \big[\textbf{X}^{(1)}, \textbf{X}^{(2)}, \ldots, \textbf{X}^{(K)}\big]$ is  the $K$ multimodal matrix of predictors, $\boldsymbol{\beta} \in \mathbb{R}^p$ the associated parameter, and $\boldsymbol{\varepsilon}$ the random error. For this purpose, we consider a data sample $\lbrace\textbf{y}=(y_i), \textbf{X}=(x_{ij})\rbrace_{1\leq i \leq n, 1\leq j \leq p}$ of size $n$ and $p$ predictors generated from some multivariate distribution with mean $\boldsymbol{0}_{p\times 1}$ and covariance matrix $\Sigma.$ Note that, each of the $K$ modalities have $p_k$ predictors for $k\in \lbrace 1, 2, \ldots, K \rbrace.$ As in high-dimensional settings, where sparsity assumptions are commonly made \citep{hastie2009elements, buhlmann2011statistics}, we assume that only a finite set of predictors contributes to the prediction of the response variable. Accordingly, to obtain a sparse estimator of the true regression coefficient $\boldsymbol{\beta}^0\in\mathbb{R}^p,$ we solve the following penalized optimization problem:
\begin{equation}\label{eqn:high_dim}
    \widehat{\beta} = \argmin_{\boldsymbol{\beta}\in \mathbb{R}^p} \frac{1}{2n} \|\textbf{y} - \textbf{X}\boldsymbol{\beta}\|^2 + \lambda \mathcal{L}(\boldsymbol{\beta}) 
\end{equation}
where $\widehat{\boldsymbol{\beta}}$ is the estimator of the true unknown parameter $\boldsymbol{\beta}^0, \ \lambda>0$ a tuning parameter and $\mathcal{L}(\cdot)$ a generic penalty function such as LASSO \citep{lasso_tib}, used to obtain a sparse solution. Note that, optimization of equation (\ref{eqn:high_dim}) is equivalent to solving: $\argmin_{\boldsymbol{\beta}\in \mathbb{R}^p} \frac{1}{2} \boldsymbol{\beta}\transpose \Sigma \boldsymbol{\beta} + \boldsymbol{\beta}\transpose\textbf{C}+ \mathcal{L}(\boldsymbol{\beta}),$ where $\Sigma  = \frac{1}{n}\textbf{X}\transpose\textbf{X}$ is the positive semi-definite covariance matrix and $\textbf{C} = \frac{1}{n}\textbf{X}\transpose\textbf{y}$ the covariance vector between the predictors and the response variable. However, in practice, the matrix $\textbf{X} = (x_{ij})_{1\leq i \leq n, 1\leq j \leq p}$ is not directly observable, especially in high dimensions, and consequently $\Sigma$ and $\textbf{C}.$ Instead, we have a multimodal matrix $\textbf{Z}  = \big[\textbf{Z}^{(1)}, \textbf{Z}^{(2)}, \ldots, \textbf{Z}^{(K)}\big] = (z_{ij})_{1\leq i \leq n, 1\leq j \leq p},$ affected simultaneously by additive measurement errors and missing data, and we need to find an unbiased estimates $\widehat{\Sigma}$ of $\Sigma$ and $\widehat{\textbf{C}}$ of $\textbf{C}$ to solve the following problem:
\begin{equation}\label{eqn:high_dim_missing_error}
    \argmin_{\boldsymbol{\beta}\in \mathbb{R}^p}  \boldsymbol{\beta}\transpose \widehat{\Sigma}\boldsymbol{\beta} + \boldsymbol{\beta}\transpose\widehat{\textbf{C}} + \mathcal{L}(\boldsymbol{\beta}).
\end{equation}
Without loss of generality, we adopt the framework of \citep{yu_optimal_2020} and assume a block-missing multimodality data. For each sample, if a certain modality has missing entries, all the observations from that modality are missing. Assuming a missing completely at random mechanism for each modality, we consider as a natural initial estimate of $\Sigma$ the covariance matrix $\widetilde{\Sigma} = \frac{1}{n}\textbf{Z}\transpose\textbf{Z} = (\widetilde{\sigma}_{jt})_{j,t=1,\ldots,p}$ where $\widetilde{\sigma}_{jt} = \frac{1}{n_{jt}}\sum_{i\in\text{S}_{jt}}z_{ij}z_{it}$ is estimated using all the available data. The set $\text{S}_{jt} = \lbrace i: z_{ij} \  \text{and} \  z_{it} \ \text{are not missing} \rbrace$ represent the set of individuals with non missing data for the pair of predictors $j$ and $t$ and $n_{jt}$ is the cardinal number of $\text{S}_{jt}.$ When $j=t$ then $\text{S}_{jt} = \text{S}_{j}$ and $n_{jt}=n_j.$ However, $\widetilde{\Sigma}$ may have negative eigenvalues due to the unequal sample sizes and is not a good estimate of the covariance matrix $\Sigma,$ and not suitable to be used in (\ref{eqn:high_dim_missing_error}) directly \citep{yu_optimal_2020}.

Following the observation of \citep{yu_optimal_2020}, we note that the covariance matrix $\widetilde{\Sigma}$ can be partitioned into the form $\widetilde{\Sigma} =  \widetilde{\Sigma}_I + \widetilde{\Sigma}_C$ where $\widetilde{\Sigma}_I = 
\diag(\widetilde{\Sigma}_{11}, \ \widetilde{\Sigma}_{22}, \ldots, \widetilde{\Sigma}_{KK}) = \diag(\widetilde{\Sigma}_{I1}, \ \widetilde{\Sigma}_{I2}, \ldots, \widetilde{\Sigma}_{IK})$ is the intra-modality sample covariance matrix, and $\widetilde{\Sigma}_C = \widetilde{\Sigma} - \widetilde{\Sigma}_I$ the cross-modality sample covariance matrix. The matrix $\widetilde{\Sigma}_C$ is an off-diagonal matrix and contains $K(K-1)$ blocks matrix denoted by $\widetilde{\Sigma}_{jt}$'s, where $j, \ t\in\lbrace 1,2,\ldots,p\rbrace$ and $\widetilde{\Sigma}_{jt}$ is a $p_j \times p_t$ matrix. We also note $\Sigma_I = \diag(\Sigma_{I1}, \ \Sigma_{I2}, \ldots, \Sigma_{IK})$  and $\Sigma_C$ the true intra-modality and cross-modality covariance matrix. Likewise, $\Sigma_{jt}$ is the true covariance matrix of $\widetilde{\Sigma}_{jt}$ for $j, \ t\in\lbrace 1,2,\ldots,p\rbrace.$ 

We now address the mitigation of measurement error and missing data issues in two sequential steps, starting with the case of additive measurement error. In the presence of additive measurement error, \citet{datta_cocolasso_2017} proposed the CocoLasso estimator $\grave{\Sigma} = \widetilde{\Sigma} + \gamma^2\mathbb{I}_p$ as an unbiased estimator of the covariance matrix, where $\gamma^2$ is the variance of the additive error matrix \citep{loh_Wainwright_2012}. By applying the CocoLasso estimator separately to the intra-modality covariance matrix $\grave{\Sigma}_I = \widetilde{\Sigma}_I + \gamma^2_I \mathbb{I}_p$ and cross-modality covariance matrix $\grave{\Sigma}_C = \widetilde{\Sigma}_C + \gamma^2_C \mathbb{I}_p,$ we can mitigate the additive measurement errors and then solve the following optimization problem
\begin{equation}\label{eqn:coco}
    \argmin_{\boldsymbol{\beta}\in \mathbb{R}^p}  \boldsymbol{\beta}\transpose [\grave{\Sigma}_I + \grave{\Sigma}_C] \boldsymbol{\beta} + \boldsymbol{\beta}\transpose\widetilde{\textbf{C}} + \mathcal{L}(\boldsymbol{\beta})
\end{equation}
where $\widetilde{\textbf{C}} = (\widetilde{\text{c}}_1, \widetilde{\text{c}}_2, \ldots, \widetilde{\text{c}}_p)\transpose$ is the estimate of the cross-covariance vector $\textbf{C}$ with $\widetilde{\text{c}}_j = \frac{1}{n_j}\sum_{i\in\text{S}_j} y_i x_{ij}.$ 

However, as the CoCoLasso covariance estimator only mitigates the influence of the additive measurement error, the estimator derived from (\ref{eqn:coco}) will still be biased due to the presence of missing data. As a second step, we combine the DISCOM estimator \citep{yu_optimal_2020} to the CocoLasso covariance estimator in order to mitigate the presence of missing data and ultimately reduce simultaneously the influence of both issues. The DISCOM covariance estimator for multimodal data is a linear combination of the intra-modality covariance matrix, the cross-modality covariance matrix and the identity matrix. Thus, the covariance estimator that takes into account both measurement errors and missing data is defined by 
\begin{equation}\label{eqn:varDiscom}
\widetilde{\widetilde{\Sigma}} = \widetilde{\alpha}_1 \grave{\Sigma}_I + \widetilde{\alpha}_2 \grave{\Sigma}_C + \widetilde{\alpha}_3  \mathbb{I}_p
\end{equation}
where $\widetilde{\alpha}_1, \ \widetilde{\alpha}_2$ et $\widetilde{\alpha}_3$ are non-random weights that can be optimized by considering all linear combinations. Considering all possible linear combinations, we can find the optimal linear combination $\widehat{\Sigma}^{*} = \alpha_1^{*} \grave{\Sigma}_I + \alpha_2^{*}\grave{\Sigma}_C + \alpha_3^{*}  \mathbb{I}_p$ which minimizes the expected quadratic loss $\E\Big[\|\widehat{\Sigma}^{*} - \Sigma\|^2_F\Big].$ Furthermore, if we consider the expressions of the intra-modality covariance and cross-modality covariance matrices in (\ref{eqn:varDiscom}), we see that
\begin{equation} \label{eqn:CocoDiscom}
\begin{split}
\widetilde{\widetilde{\Sigma}} & = \widetilde{\alpha}_1 (\widetilde{\Sigma}_I + \gamma^2_I \mathbb{I}) + \widetilde{\alpha}_2 (\widetilde{\Sigma}_C + \gamma^2_C \mathbb{I}) + \widetilde{\alpha}_3  \mathbb{I}_p \\
 & = \alpha_I \widetilde{\Sigma}_I + \alpha_C \widetilde{\Sigma}_C +  \alpha_p \mathbb{I}_p
\end{split}
\end{equation}
where $\alpha_I = \widetilde{\alpha}_1, \ \alpha_C=\widetilde{\alpha}_2$ and $\alpha_p = \widetilde{\alpha}_1 \gamma^2_I + \widetilde{\alpha}_2 \gamma^2_C +  \widetilde{\alpha}_3.$ Thus, we show that the DISCOM covariance estimator accounts for both missing data and measurement errors. These results are confirmed by the simulation results. 

\subsection{Adaptive direct sparse regression procedure using covariance from multimodality data (AdapDISCOM)}\label{ada_discom_metho}
To better capture the intrinsic heterogeneity across modalities, we introduce AdapDISCOM, an adaptive extension of the DISCOM framework. In this setting, we explicitly recognize that data from each modality are generated through distinct underlying processes. Beyond addressing the structural heterogeneity characteristic of multimodal data, AdapDISCOM also accounts for the variability in measurement errors affecting each modality. Specifically, the nature and magnitude of measurement errors differ substantially depending on the acquisition technique: for example, errors resulting from genetic sequencing are fundamentally different from those arising in medical imaging, as well as from inaccuracies associated with self-reported clinical or environmental measures. AdapDISCOM extends DISCOM by introducing a modality-specific covariance estimator, whose parameters are adapted to the characteristics of each modality, thereby enabling a more accurate modeling of the cross-modal variability. We therefore propose to employ the following estimator:
\begin{equation} 
\widehat{\Sigma} = \alpha_1 \widetilde{\Sigma}_{I_1} + \alpha_2 \widetilde{\Sigma}_{I_2} + \cdots + \alpha_K \widetilde{\Sigma}_{I_K} + \alpha_C \widetilde{\Sigma}_{C} +  \alpha_p \mathbb{I}_p,
\end{equation}
where $\alpha_k, \ k=1,2,\ldots, K, \ \alpha_C,$ and $\alpha_p$ are nonrandom weights. Considering all possible linear combinations, we can find the optimal linear combination $\widehat{\Sigma}^{*} = \alpha_1^{*} \widetilde{\Sigma}_{I_1} + \alpha_2^{*} \widetilde{\Sigma}_{I_2} + \cdots + \alpha_K^{*} \widetilde{\Sigma}_{I_K}+ \alpha_C^{*} \widetilde{\Sigma}_{C} +  \alpha_p^{*} \mathbb{I}_p$ whose expected quadratic loss $\E\Big[\|\widehat{\Sigma}^{*} - \Sigma\|^2_F\Big]$ is the minimum. The optimal weight are shown in the following proposition.
\begin{proposition} \label{prop1}
Consider the following optimization problem
\begin{equation*}
    \min_{\alpha_k,\alpha_C, \alpha_p}\E\Big[\|\widehat{\Sigma} - \Sigma\|^2_F\Big] \; \ \text{subject to} \ \widehat{\Sigma} = \alpha_1 \widetilde{\Sigma}_{I_1} + \alpha_2 \widetilde{\Sigma}_{I_2} + \cdots + \alpha_K \widetilde{\Sigma}_{I_K} + \alpha_C \widetilde{\Sigma}_{C} +  \alpha_p \mathbb{I}_p,
\end{equation*}    
where the weights $\alpha_k, \ k=1,2,\ldots, K, \ \alpha_C,$ and $\alpha_p$ are nonrandom. Denote for, $k=1,2,\ldots, K, \ \delta_{C}^2 = \E[\|\widetilde{\Sigma}_{C} - \Sigma_{C}\|^2_F], \ \delta_{I_k}^2 = \E[\|\widetilde{\Sigma}_{I_k} - \Sigma_{I_k}\|^2_F]$ and $\theta_{I_k}^2 = \E[\|\gamma^*\mathbb{I}_p - \Sigma_{I_k}\|^2_F]$ where
\begin{equation*}
\begin{split}
    \gamma^* & = \frac{1}{p}\times \frac{(1-\alpha_1^*)^2\Tr(\widetilde{\Sigma}_{I_1}) + (1-\alpha_2^*)^2\Tr(\widetilde{\Sigma}_{I_2}) + \cdots + (1-\alpha_K^*)^2\Tr(\widetilde{\Sigma}_{I_K})}{(1-\alpha_1^*)^2 + (1-\alpha_2^*)^2 + \cdots + (1-\alpha_K^*)^2} \\
             & = \frac{1}{p} \sum_{k=1}^{K} \frac{(1- \alpha_k^*)^2}{\sum_{t=1}^{K}(1-\alpha_t^*)^2} \Tr(\widetilde{\Sigma}_{I_k}).
\end{split}
\end{equation*}
The optimal weights are, for $k=1,2,\ldots, K,$
\begin{align*} 
& \alpha_k^*  = \frac{\theta_{I_k}^2}{\theta_{I_k}^2 + \delta_{I_k}^2} \in[0,1], \quad\quad \alpha_C^* = \frac{\|\Sigma_{C}\|^2_F}{\|\Sigma_{C}\|^2_F + \delta_{C}^2} \in[0,1] \quad \text{and} \\ 
 & \alpha_p^*=\gamma^*(K - \alpha_1^* - \alpha_2^* - \ldots - \alpha_K^*) = \gamma^* \sum_{k=1}^K (1 - \alpha_k^*).
\end{align*}
In addition, we have
\begin{equation*}
    \begin{split}
        \E\Big[\|\widetilde{\Sigma}^* - \Sigma\|^2_F\Big] & = \sum_{k=1}^K\frac{\delta_{I_k}^2\theta_{I_k}^2}{\theta_{I_k}^2 + \delta_{I_k}^2} + \frac{\delta_{C}^2\|\Sigma_{C}\|^2_F}{\delta_{C}^2 + \|\Sigma_{C}\|^2_F} \leq \sum_{k=1}^K \delta_{I_k}^2 + \delta_{C}^2 \\
        & = \E\Big[\|\widetilde{\Sigma} - \Sigma\|^2_F\Big].
    \end{split}
\end{equation*}
\end{proposition}
Note that, the optimal parameters of AdapDISCOM are a generalization of those of DISCOM and when $K=1,$ or when all parameters $\alpha_k^*$ are equal then Proposition \ref{prop1} is the same as Proposition 1 shown in \citep{yu_optimal_2020}. Furthermore, the expression of $\gamma^*$ shows how AdapDISCOM accounts for the heterogeneity of each modality's errors, where  the estimator of the intra-modality covariance matrix of modality $k$ can be considered to be  $\alpha_k^*\widetilde{\Sigma}_{I_k} + w_k \Tr(\widetilde{\Sigma}_{I_1}/p)\mathbb{I}_p,$ where $w_k = \frac{(1-\alpha_k^*)^2}{\sum_{k=1}^{K}(1-\alpha_k^*)^2}.$ 

The relative improvement in the expected quadratic loss over the sample covariance matrix is equal to 
\begin{equation*}
    \begin{split}
        \frac{\E\Big[\|\widetilde{\Sigma} - \Sigma\|^2_F\Big] - \E\Big[\|\widetilde{\Sigma}^* - \Sigma\|^2_F\Big] }{\E\Big[\|\widetilde{\Sigma} - \Sigma\|^2_F\Big]} & = \sum_{k=1}^K\frac{\delta_{I_k}^2}{\sum_{t=1}^K\delta_{I_t}^2 + \delta_{C}^2} \cdot (1-\alpha_k^*) \\
        & + \frac{\delta_{C}^2}{\sum_{t=1}^K\delta_{I_t}^2 + \delta_{C}^2} \cdot (1-\alpha_C^*).
    \end{split}
\end{equation*}
Thus, when a group of modalities is relatively less contaminated by measurement errors and missing data, then their intra-modality covariance matrix is relatively precise $(\delta_{I_k}^2 \ \text{is small})$ and their optimal weight $\alpha_k^*  = \frac{\theta_{I_k}^2}{\theta_{I_k}^2 + \delta_{I_k}^2}$ should be large compared to other modalities, and the percentage of improvement is relatively small. This highlights the importance of using weights adapted to each modality, given that some modalities may be less contaminated than others and therefore have a more precise covariance. In the same way, the intra-modality covariance matrix is more accurate than the inter-modality covariance matrix, due to the unequal sample sizes \citep{yu_optimal_2020}.

$\textbf{Robust estimates of} \ \Sigma \ \text{ and } \textbf{C} $ 

Similar to DISCOM, we propose a robust estimators of $\Sigma$ and $\text{C}$ in the case of heavy-tailed distributions, where $\widetilde{\Sigma}$ and $\widetilde{\text{C}}$ estimators may perform poorly. We construct a robust initial estimates $\breve{\Sigma} = (\breve{\sigma}_{jt})_{j,t=1,2,\ldots,p}$ of $\Sigma$ and $\breve{C}=(\breve{c}_1,\breve{c}_2, \ldots, \breve{c}_p)\transpose$ of $C$ given by
\begin{equation*}
\begin{split}
    \breve{\sigma}_{jt} & = \ \text{the solution to} \quad \sum_{i\in S_{jt}}\psi_{H_{jt}}(x_{ij}x_{it} - \mu) = 0 \quad\quad \text{and} \\
    \breve{c}_j & = \ \text{the solution to} \quad  \sum_{i\in S_{j}}\psi_{H_{j}}(x_{ij}y_i - \mu) = 0,
\end{split}
\end{equation*}
where $\psi_{H} (z) = z \cdot \mathds{1}_{\lbrace|z|\leq H\rbrace} + H \cdot \sign(z) \mathds{1}_{\lbrace|z|> H\rbrace}$ is the Huber function \citep{huber1964}, $\mu$ the mean and $H$ an hyperparameter with default value  chosen to be $1.345.$ However, in this setting we propose to use flexibly different values of $H$ to account for the different numbers of samples available. The estimator built from the initial robust estimators will be designated AdapDISCOM-Huber in the following. 

\subsection{Theoretical Study}\label{theo_study}
In this section, we provide the theoretical properties of the proposed estimators. We begin by establishing their convergence properties under the assumption that both the response variable and the predictor matrix follow sub-Gaussian distributions. We then extend these results to the case where the response and predictors exhibit heavy-tailed distributions. Next, we demonstrate the model selection consistency of our proposed method and conclude the section with the derivation of the estimator in the context of optimal linear prediction.

As in the original DISCOM framework, we assume in this section that the true variances of all predictors are equal to 1. Similarly, the variance estimators $\widetilde{\sigma}_{jj},$  including the Huber-based estimators $\breve{\sigma}_{jj}$, are scaled to 1, for each $j\in\lbrace 1, 2, \ldots, p \rbrace.$ We further assume that the parameter vector $\boldsymbol{\beta}^0$ is sparse, with $J = \lbrace j : \beta_j^0 \ne 0 \rbrace$ denoting the index set of the important predictors and $s =|J|$ its cardinality. Finally, we define $\beta_{\max}^0 = \max_{j\in J}|\beta_j^0|$ and $\beta_{\min}^0 = \min_{j\in J}|\beta_j^0|.$

\subsubsection{Sub-Gaussian Case}\label{conv_rate1}
The results in this subsection are based on the following assumptions.

$(\textbf{A1}).$ Suppose that there exists a constant $L>0$ such that
\begin{align*}
 &  \E[\exp ( t \boldsymbol{x}_j)]  \leq \exp \Bigg( \frac{L^2 t^2}{2}\Bigg) \; \text{\ for all \ } j\in \lbrace 1, 2, \ldots, p\rbrace \text{ \ and \ } t \in \mathbb{R}, \\
  & \E[\exp ( t \boldsymbol{y})] \leq \exp \Bigg( \frac{L^2 t^2}{2}\Bigg) \; \text{\ for all \ } t \in \mathbb{R}.
\end{align*}
$(\textbf{A2}).$ Suppose that the true covariance matrix $\Sigma$ satisfies the following restricted eigenvalue (RE) condition:
\begin{equation*}
   \min_{\boldsymbol{\delta}\in\lbrace \boldsymbol{u} \in \mathbb{R}^p: \| \boldsymbol{u}_{J^c}\|_1\leq 7 \| \boldsymbol{u}_{J}\|_1  \rbrace} \frac{\boldsymbol{\delta}\transpose\Sigma\boldsymbol{\delta}}{\boldsymbol{\delta}\transpose \boldsymbol{\delta}} \ge m \ge 0.
\end{equation*}
Condition \textbf{A1} posits that both the predictors and the response variable follow sub-Gaussian distributions, while the RE condition \textbf{A2} is used to derive statistical error bounds for the Lasso estimate \citep{datta_cocolasso_2017}. The following theorem shows the large deviation bounds of $\widetilde{\Sigma}$ and $\widetilde{\boldsymbol{C}}$ as well as the convergence rate of $\| \widetilde{\boldsymbol{\beta}} - \boldsymbol{\beta}^0\|_2.$
\begin{theorem} \label{theo 1}~~\\
\textbf{Convergence rate of} \; $\widetilde{\Sigma}.$ \; Under condition $(\textbf{A1}),$ if $\min_{j,t,k} n_{jt}^k \geq 6 \log p,$ there exists two positive constants $\nu_1 = 8 \sqrt{6}(1 + 4\text{L}^2)$ and $\nu_2 = 4$ such that
\begin{equation*}
    \begin{split}
        & \max_{j,t,k} P \Bigg( |\widetilde{\sigma}_{jt}^k - \sigma_{jt}^k| \geq \nu_1 \sqrt{\frac{\log p}{n_{jt}^k}}\Bigg) \leq \frac{\nu_2}{p^3}, \\
        & P \Bigg( \| \widetilde{\Sigma} - \Sigma \|_{\max} \geq \nu_1 \sqrt{\frac{\log p}{\min_{j,t,k} n_{jt}^k}} \Bigg) \leq \frac{\nu_2}{p}. 
    \end{split}
\end{equation*}
\textbf{Convergence rate of} \; $\widetilde{\text{C}}.$ \;  Given the following constants 
$\nu_3 = 16 (1 + 4\frac{\text{L}^2}{\min\lbrace\Var(y), 1\rbrace} )$ and $\nu_4 = 4$ we have 
\begin{equation*}
    \begin{split}
        & \max_{j,k} P \Bigg( |\widetilde{\text{c}}_{j}^k - \text{c}_{j}^k| \geq \nu_3 \sqrt{\frac{\log p}{n_{j}^k}}\Bigg) \leq \frac{\nu_4}{p^2}, \\
        & P \Bigg( \| \widetilde{\text{C}} - \text{C} \|_{\max} \geq \nu_3 \sqrt{\frac{\log p}{\min_{j,k} n_{j}^k}} \Bigg) \leq \frac{\nu_4}{p}. 
    \end{split}
\end{equation*}
$\textbf{Convergence rate of}$  \; $\widetilde{\boldsymbol{\beta}}.$ \; Let $1-\alpha_k = \bigO\Big(\sqrt{(\log  p/\min_j n_j^k)}\Big)$ and $1-\alpha_C= \bigO\Big(\sqrt{(\log  p/\min_{j,t,k} n_{jt}^k)}\Big).$ With the convergence rates of $\ \widetilde{\Sigma} \ $ and $\ \widetilde{\text{C}}\ $ and under the conditions $(\textbf{A1})$ and $ \ (\textbf{A2}), \ $ if
$\ s \sqrt{(\log p) / \min_{j,t,k} n_{jt}^k} = \smallO (1)\ $ and we choose $\ \lambda = 2 \|\widetilde{\text{C}} - \widehat{\Sigma}\boldsymbol{\beta}^0\|_{\max},\ $ then we have
\begin{equation*}
    \| \widetilde{\boldsymbol{\beta}} - \boldsymbol{\beta}^0 \|_2 = \bigO_p(\sqrt{s}\lambda) = \bigO_p\bigg(
\|\boldsymbol{\beta}^0 \|_1 \sqrt{s(\log p)/\min_{j,t,k} n_{jt}^k} \bigg).
\end{equation*}
\end{theorem}

\begin{remark}
Regarding the convergence rate of $\widetilde{\Sigma}$ and $\widetilde{\text{C}},$ Theorem \ref{theo 1} shows that
~~\\
$\| \widetilde{\Sigma} - \Sigma \|_{\max} = \bigO_p \Big(\sqrt{\log p / \min_{j,t,k} n_{jt}^k} \Big), \text{ \ and \ } \|\widetilde{\text{C}} - \text{C}\|_{\max} = \bigO_p \Big(\sqrt{\log p / \min_{j,k} n_{j}^k}\Big).$ That is, on the worst case, the performance of $\widetilde{\Sigma}$ and $\widetilde{\text{C}}$ depend when there are only $\min_{j,t,k} n_{jt}^k$ samples to estimate some entries in $\Sigma$ and $\min_{j,k} n_{j}^k$ samples to estimate the covariance between some predictors and the response variable, respectively. Moreover, the same results hold when using $n_{\text{complete}},$ the number of samples with complete observations, which can be much smaller than $\min_{j,t,k} n_{jt}^k$ and $\min_{j,k} n_{j}^k$ in the context of block-missing multimodality data. Thus, Theorem \ref{theo 1} demonstrates that all available data can be fully exploited for the first step of our proposed AdapDISCOM method.

With respect to the convergence of $\widetilde{\boldsymbol{\beta}},$ it can be shown that, under the assumptions of no missing data, Gaussian-distributed predictors, and independent and identically distributed Gaussian errors, we have from Theorem \ref{theo 1} that $\|\widetilde{\text{C}} - \widehat{\Sigma}\boldsymbol{\beta}^0\|_{\max} = \|\boldsymbol{X}\transpose\boldsymbol{\varepsilon}/n \|_{\max} = \bigO_p(\sqrt{(\log p)/n)},$ and $\| \widetilde{\boldsymbol{\beta}} - \boldsymbol{\beta}^0 \|_2 = \bigO_p(\sqrt{(s \log p)/n)},$ which is the minimax $l_2$-norm rate as shown by \citet{raskutti_2011}. Since the complete data generated from the Gaussian random design can be viewed as a special type of block-missing multimodality data, the error bound in Theorem \ref{theo 1} is sharp.

In general, the RE condition is satisfied when the design matrix \textbf{X} is fixed and complete. However, when this is not the case, for example, when the design matrix is Gaussian or random, additional assumptions are required to ensure the RE condition holds. For instance, in the case of a random design matrix, \citet{van_de_Geer_2009} showed that the empirical covariance matrix $\widehat{\Sigma} = \textbf{X}\transpose \textbf{X}/n$ satisfies the RE condition as long as the true covariance matrix $\Sigma$ does, and $s^2 \log p /  n = \smallO(1).$ Similarly, \citet{raskutti_2011} demonstrated that, for a Gaussian random design matrix,  $\widehat{\Sigma}$ satisfies the RE condition with high probability provided that the true Gaussian covariance matrix $\Sigma$ satisfies the RE condition and $n>\text{Constant} \cdot s \log p.$ Notably, their analysis is global, focusing on the full random matrix $\widehat{\Sigma} = \textbf{X}\transpose \textbf{X}/n$  rather than on individual entries of $\widehat{\Sigma}.$

In our work, as in DISCOM \citep{yu_optimal_2020}, we consider a general random design matrix encompassing both sub-Gaussian and heavy-tailed distributions, and we study a proposed estimated covariance matrix $\widehat{\Sigma} \ne \textbf{X}\transpose\textbf{X}/n$ in most cases. To ensure the RE condition holds with high probability, we adopt the assumption $\ s \sqrt{(\log p) / \min_{j,t,k} n_{jt}^k} = \smallO (1), \ $ which closely parallels the condition $\ s^2 \log p / n = \smallO (1), \ $ used by \citet{van_de_Geer_2009} for the complete data.
\end{remark}

\subsubsection{Heavy-tailed Case}\label{conv_rate2}
The results in the heavy-tailed setting are derived under the following moment condition.

(\textbf{A3}). Suppose that $\ \max_{1\leq j \leq p} \E[\boldsymbol{x}_j^4] \leq Q_1^2/48 \ $ and $ \ \E[\boldsymbol{y}^4] \leq Q_2^2, \ $ where $Q_1$ and $Q_2$ are two positive constants.

In the Lasso literature, most studies assume a fixed design matrix \citep{Meinshausen2006, zhao06a, Zou01122006} and Gaussian errors \citep{Zhang_Huang2008}, or at most sub-exponential tails \citep{bunea2008consistent, Meinshausen_Yu2009}. In contrast, following the DISCOM approach, we consider a random design matrix and assume that the fourth moments of both the predictors and the response variable are finite. Note that under condition (\textbf{A3}) the tails of the distributions of $\boldsymbol{x}_j$'s and $\boldsymbol{y}$ may not be exponentially bounded.

\begin{theorem}\label{theo2} ~~\\
\textbf{Convergence rate of} \; $\breve{\Sigma}.$ \; Under condition $(\textbf{A3}),$ let $\text{H}_{jt}^k = \frac{\text{Q}_1}{12}\sqrt{n_{jt}^k/\log p}$ for each $j,t\in\lbrace 1,2,\ldots, p\rbrace, \ k\in\lbrace 1,2,\ldots, K\rbrace,$ if $\min_{j,t,k}n_{jt}^k \geq 24 \log p,$ we have
    \begin{equation*}
    \begin{split}
        & \max_{j,t,k} P \Bigg( |\breve{\sigma}_{jt}^k - \sigma_{jt}^k| \geq \text{Q}_1 \sqrt{\frac{\log p}{n_{jt}^k}}\Bigg) \leq \frac{2}{p^3}, \\
        & P \Bigg( \| \breve{\Sigma} - \Sigma \|_{\max} \geq \text{Q}_1 \sqrt{\frac{\log p}{\min_{j,t,k} n_{jt}^k}} \Bigg) \leq \frac{2}{p}. 
    \end{split}
\end{equation*}
\textbf{Convergence rate of} \; $\breve{\text{C}}.$ \; Furthermore, let $\text{H}_j^k = (\text{Q}_1 + \text{Q}_2) \sqrt{n_j^k/\log p}$ for each $j \in \lbrace 1,2, \ldots, p \rbrace, \  k\in\lbrace 1, 2, \ldots, K\rbrace,$ we have 
\begin{equation*}
    \begin{split}
        & \max_{j,k} P \Bigg( |\breve{\text{c}}_{j}^k - \text{c}_{j}^k| \geq 8 (\text{Q}_1 + \text{Q}_2) \sqrt{\frac{\log p}{n_{j}^k}} \Bigg) \leq \frac{2}{p^2}, \\
        & P \Bigg( \| \breve{\text{C}} - \text{C} \|_{\max} \geq 8 (\text{Q}_1 + \text{Q}_2)  \sqrt{\frac{\log p}{\min_{j,k} n_{j}^k}} \Bigg) \leq \frac{2}{p}. 
    \end{split}
\end{equation*}
\textbf{Convergence rate of}  \; $\breve{\boldsymbol{\beta}}.$ \; Assuming the above results, let $1-\alpha_k = \bigO\Big(\sqrt{(\log  p/\min_j n_j^k)}\Big)$ and $1-\alpha_C= \bigO\Big(\sqrt{(\log  p/\min_{j,t,k} n_{jt}^k)}\Big),$ \, then, under the conditions $(\textbf{A2})$ and $(\textbf{A3}),$ \; if \; 
$s \sqrt{(\log p) / \min_{j,t,k} n_{jt}^k} = \smallO (1)$ and $\lambda = 2 \|\breve{\text{C}} - 
 \widehat{\Sigma}\boldsymbol{\beta}^0\|_{\max},$ we have
\begin{equation*}
    \| \breve{\boldsymbol{\beta}} - \boldsymbol{\beta}^0 \|_2 = \bigO_p(\sqrt{s}\lambda) = \bigO_p\bigg(
\|\boldsymbol{\beta}^0 \|_1 \sqrt{s(\log p)/\min_{j,t,k} n_{jt}^k} \bigg).
\end{equation*}
\end{theorem}
\begin{remark}
Theorem \ref{theo2} establishes that, under assumption (\textbf{A3}), the convergence rate of $\| \breve{\boldsymbol{\beta}} - \boldsymbol{\beta}^0 \|_2$ in the heavy-tailed setting matches that of $\| \widetilde{\boldsymbol{\beta}} - \boldsymbol{\beta}^0 \|_2$   in the sub-Gaussian setting. Consistent with the findings in DISCOM \cite{yu_optimal_2020}, our Huber’s M-estimators in the heavy-tailed setting achieve the same convergence rate as the sample covariance-based estimators in the sub-Gaussian case. Furthermore, when the predictor and response distributions do not have exponentially bounded tails, the large deviation bounds for  $ \ \widetilde{\Sigma} \ $ and $ \ \widetilde{C} \ $ can be wider than those of the Huber's M-estimators $ \ \breve{\Sigma} \ $ and $ \ \breve{C}, \ $ respectively. For additional details, we refer the reader to \citep{yu_optimal_2020}.

Similarly, when $p$ is fixed, comparable bounds can be obtained in the sub-Gaussian setting. Moreover, as shown by \citet{yu_optimal_2020}, the convergence rate of the estimation error in the classical fixed-$p$ regime is significantly faster than in the high-dimensional setting where $p \longrightarrow \infty.$
\end{remark}
 
\subsubsection{Model Selection Consistency}\label{consis_theo}
The property of model selection consistency for the AdapDISCOM method holds under the following assumption.

(\textbf{A4}). $\|\Sigma_{J^cJ}\Sigma_{JJ}^{-1}\|_{\infty} \leq 1 - \eta,$ where $\eta\in (0,1)$ is a constant, $\Sigma_{J^cJ}$ is the sub-matrix of $\Sigma$ with row indices in the set $J^c$ and column indices in the set $J,$ and $\Sigma_{JJ}$ is the sub-matrix of $\Sigma$ with both row and column indices in the set $J.$

According to \citet{yu_optimal_2020}, condition (\textbf{A4}) can be viewed as the population-level counterpart of the strong irrepresentable condition proposed in \citet{zhao06a}. The following theorem show the model selection consistent for the AdapDISCOM method in the sub-Gaussian and heavy-tailed case.

\begin{theorem} ~~\\
\textbf{Sub-Gaussian Case}.  \; Under conditions $(\textbf{A1})$ and $(\textbf{A4}),$ let $1-\alpha_k = \bigO\Big(\sqrt{(\log  p/\min_j n_j^k)}\Big)$ and $1-\alpha_C= \bigO\Big(\sqrt{(\log  p/\min_{j,t,k} n_{jt}^k)}\Big),$ \, If
\begin{equation*}
\|(\Sigma_{JJ})^{-1}\|_{\infty} \cdot \sqrt{\frac{s^2\log p}{\min_{j,t,k} n_{jt}^k}}  \longrightarrow 0, \;  
\frac{1+s\boldsymbol{\beta}_{\max}^0}{\lambda} \sqrt{\frac{\log p}{\min_{j,t,k} n_{jt}^k}} \longrightarrow 0, \;
\text{and} \; \frac{\lambda\cdot\|(\Sigma_{JJ})^{-1}\|_{\infty}}{\boldsymbol{\beta}_{\min}^0} \longrightarrow 0,
\end{equation*}
then there exists a solution $\widetilde{\boldsymbol{\beta}}$ such that 
\begin{equation*}
    P(\sign(\widetilde{\boldsymbol{\beta}}) = \sign(\boldsymbol{\beta}^0)) \longrightarrow 1, \; \text{as} \; \min_{jtk}n_{jt}^k \longrightarrow \infty \; \text{and} \; p \longrightarrow \infty.
\end{equation*}
\textbf{Heavy-Tailed Case}.  \; Under conditions $(\textbf{A3})$ and $(\textbf{A4}),$ let $1-\alpha_k = \bigO\Big(\sqrt{(\log  p/\min_j n_j^k)}\Big)$ and $1-\alpha_C= \bigO\Big(\sqrt{(\log  p/\min_{j,t,k} n_{jt}^k)}\Big), \; \text{H}_{jt}^k = \frac{\text{Q}_1}{12}\sqrt{n_{jt}^k/\log p}$ and \; $\text{H}_j^k = (\text{Q}_1 + \text{Q}_2) \sqrt{n_j^k/\log p}.$ \, If 
\begin{equation*}
\|(\Sigma_{JJ})^{-1}\|_{\infty} \cdot \sqrt{\frac{s^2\log p}{\min_{j,t,k} n_{jt}^k}}  \longrightarrow 0, \;  
\frac{1+s\boldsymbol{\beta}_{\max}^0}{\lambda} \sqrt{\frac{\log p}{\min_{j,t,k} n_{jt}^k}} \longrightarrow 0, \;
\text{and} \; \frac{\lambda\cdot\|(\Sigma_{JJ})^{-1}\|_{\infty}}{\boldsymbol{\beta}_{\min}^0} \longrightarrow 0,
\end{equation*}
then there exists a solution $\breve{\boldsymbol{\beta}}$ such that  
 \begin{equation*}
    P(\sign(\breve{\boldsymbol{\beta}}) = \sign(\boldsymbol{\beta}^0)) \longrightarrow 1, \; \text{as} \; \min_{jtk}n_{jt}^k \longrightarrow \infty \; \text{and} \; p \longrightarrow \infty.
\end{equation*}
\end{theorem}

\begin{remark}
In the specific case of a Gaussian design, \citet{wainwright2009sharp} showed that, by leveraging concentration inequalities for the normal distribution and the fact that $\widehat{\Sigma} = \textbf{X}\transpose\textbf{X}/n$ for the complete data, one can establish model selection consistency provided $n > \text{Constant} \cdot s \log (p-s).$ In our setting, however, we consider a general random design encompassing both sub-Gaussian and heavy-tailed distributions, and $\widehat{\Sigma} \ne \textbf{X}\transpose\textbf{X}/n$ due to the block-wise missingness in multimodal data. Therefore, we rely on $\|(\Sigma_{JJ})^{-1}\|_{\infty} \cdot \sqrt{\frac{s^2\log p}{\min_{j,t,k} n_{jt}^k}} = \smallO (1)$ to ensure that properties $\|(\widehat{\Sigma}_{JJ})^{-1}\|_{\infty} \leq \text{Constant}\cdot \|(\Sigma_{JJ})^{-1}\|_{\infty}$ and $\|\widehat{\Sigma}_{J^cJ} \widehat{\Sigma}_{JJ}^{-1} \|_{\infty} \leq 1-\eta\prime$ are satisfied with high probability, if $\|\Sigma_{J^cJ}\Sigma_{JJ}^{-1}\|_{\infty} \leq 1-\eta.$ The parameters $\eta\prime \in (0, 1)$ and $\eta \in (0, 1)$ are two constants. It is worth noting that this condition has also been used to establish model selection consistency for random designs, as in \citep{jeng2011sparse}.
\end{remark}
\subsubsection{Efficient estimation in the Optimal Linear Prediction}\label{fast-estimate}
Given initial estimates of $\Sigma$ and $\textbf{C},$ for example $\widetilde{\Sigma}=\sum_{k=1}^{K} \widetilde{\Sigma}_{I_k} + \widetilde{\Sigma}_{C}$ and $\widetilde{\textbf{C}},$ the proposed AdapDISCOM method estimates $\boldsymbol{\beta}^0$ by solving the following optimization problem:
\begin{equation}\label{main_adapDISCOM}
\min_{\boldsymbol{\beta}} \frac{1}{2} \boldsymbol{\beta}\transpose \bigg[\sum_{k=1}^{K}\alpha_k \widetilde{\Sigma}_{I_k}  + \alpha_C \widetilde{\Sigma}_{C} +  \sum_{t=1}^{K} \sum_{k=1}^{K} (1-\alpha_t)\frac{(1-\alpha_k)^2}{\sum_{t=1}^{K}(1-\alpha_k)^2} \frac{\Tr(\widetilde{\Sigma}_{I_k})}{p}\mathbb{I}_p \bigg] \boldsymbol{\beta} - \widetilde{\textbf{C}}\transpose \boldsymbol{\beta} + \lambda \| \boldsymbol{\beta} \|_1,    
\end{equation}
where $\lbrace\alpha_k\rbrace_{k=1,2,\ldots, K} \in (0, 1), \ \alpha_C \in (0, 1)$ are weights and $\lambda > 0$ the tuning parameter.

In practice, the various hyperparameters, $\lbrace\alpha_k\rbrace_{k=1,2,\ldots, K}, \ \alpha_C$ and $\lambda$ can be chosen through application of cross-validation selection method. Alongside this approach, we present an efficient tuning procedure which builds on our theoretical results. The idea is to choose reasonable $\alpha_k \in [0,1]$ for $k=1,2,\ldots,K$ and $\alpha_C \in [0,1]$ to guarantee that the estimated covariance matrix $\widehat{\Sigma}$ is positive semi-definite. We can achieve that by considering $\alpha_k$ and $\alpha_C$ such that the smallest eigen value is positive, $\kappa(\widehat{\Sigma}) \geq 0.$ 

From the previous theoretical results, we have $1-\alpha_k = \bigO\Big(\sqrt{(\log  p/\min_j n_j^k)}\Big)$ and $1-\alpha_C= \bigO\Big(\sqrt{(\log  p/\min_{j,t} n_{jt})}\Big),$ let $m_k = \sqrt{\frac{\log  p}{\min_j n_j^k }}$ and $m_C = \sqrt{\frac{\log  p}{\min_{j,t} n_{jt}}}.$ Then set, $\alpha_k=1-l_0m_k$ and $\alpha_C=1-l_0m_C,$ where $l_0\in[l_{\min},l_{\max}]$ is a tuning parameter. From the definition, we have that $l_0$ should satisfy $l_0>0.$ To guarantee that both $\alpha_k$ and $\alpha_C$ are nonnegative,  we set $l_{\max} = \min\big(\min_k\frac{1}{m_k}, \frac{1}{m_C}\big) = \frac{1}{m_C}$   since $m_C > m_k,$ for $k\in\lbrace 1,2,\ldots, K\rbrace.$

Now consider 
\begin{equation*}
    \widehat{\Sigma} = \widetilde{\Sigma} - l_0m_C\widetilde{\Sigma} + l_0\sum_{k=1}^{K} (m_C-m_k)\widetilde{\Sigma}_{I_k} + \frac{l_0}{p}\sum_{t=1}^{K} \sum_{k=1}^{K} m_t\bigg(\frac{m^2_k\Tr(\widetilde{\Sigma}_{I_k})}{\sum_{t=1}^{K}m_t^2} \bigg)\mathbb{I}_p
\end{equation*}
then, having $\kappa_{\min}(\widehat{\Sigma}) \geq 0 $ is equivalent to
\begin{equation*}
    l_0 \geq \frac{-\kappa_{\min}(\widetilde{\Sigma})}{-m_C\kappa_{\min}(\widetilde{\Sigma}) + \kappa_{\min} \bigg( \sum_{k=1}^{K}(m_C-m_k)\widetilde{\Sigma}_{I_k} + \frac{1}{p}\sum_{t=1}^{K}\sum_{k=1}^{K} m_t\bigg(\frac{m^2_k\Tr(\widetilde{\Sigma}_{I_k})}{\sum_{t=1}^{K}m_t^2} \bigg)\mathbb{I}_p \bigg)}.
\end{equation*}

To ensure that $l_0$ satisfies this condition, we'll specify its lower bound $l_{\min}.$ Therefore, if $\kappa_{\min}(\widetilde{\Sigma})\geq 0,$ we choose $l_{\min}=0$ to guarantee that $\kappa_{\min}(\widehat{\Sigma}) \geq 0.$ Otherwise, we choose
\begin{equation*}
    l_{\min} = \frac{-\kappa_{\min}(\widetilde{\Sigma})}{-m_C\kappa_{\min}(\widetilde{\Sigma}) + \kappa_{\min} \bigg( \sum_{k=1}^{K}(m_C-m_k)\widetilde{\Sigma}_{I_k} + \frac{1}{p}\sum_{t=1}^{K}\sum_{k=1}^{K} m_t\bigg(\frac{m^2_k\Tr(\widetilde{\Sigma}_{I_k})}{\sum_{t=1}^{K}m_t^2} \bigg)\mathbb{I}_p \bigg)},
\end{equation*}
and given that $m_C\geq m_k > 0$ and
\begin{equation}\label{fast_tuning}
    \sum_{k=1}^{K}(m_C-m_k)\widetilde{\Sigma}_{I_k} + \frac{1}{p}\sum_{t=1}^{K} \sum_{k=1}^{K} m_t \bigg(\frac{m^2_k\Tr(\widetilde{\Sigma}_{I_k})}{\sum_{t=1}^{K}m_t^2} \bigg)\mathbb{I}_p \; \text{ is positive definite},
\end{equation}
 we have $l_{\min}$ always less than $l_{\max} = 1/m_C.$ Therefore, instead of having to estimate $K+2$ hyperparameters, this parametrization reduces the tuning to the search for the best $l_0 \in [l_{\min}, l_{\max}]$ and the parameter $\lambda.$ Furthermore, rather than performing the eigendecomposition for each parameter combination to ensure that $\widehat{\Sigma}$ is positive semidefinite, this fast reparametrization method requires two eigendecomposition of $\widetilde{\Sigma}$ and the matrix from (\ref{fast_tuning}) before the tuning parameter selection process. Similar to \citep{yu_optimal_2020}, for each $l_0 \in [l_{\min}, l_{\max}],$ we can incorporate the coordinate descent algorithm \citep{friedman2010regularization} on a grid of $\lambda$ values or apply the LARS algorithm \citep{jeng2011sparse} to compute the solution path.  

The resulting method from this procedure is referred to as FastAdapDISCOM. Note that, when $k=1$ or $m_k =  m$ for all $k \in \lbrace 1, 2, \ldots, K \rbrace$ then  FastAdapDISCOM reduces to the FastDISCOM method proposed by \citep{yu_optimal_2020}.

\section{Simulation}
In this section, we conduct simulation study to assess the performance of  the AdapDISCOM family methods within the linear model framework $\textbf{y} = \textbf{X}\boldsymbol{\beta}^0 + \boldsymbol{\varepsilon}.$ We refer to our proposed methods based on the sample covariance estimates and Huber’s M-estimates as AdapDISCOM and AdapDISCOM-Huber, respectively. When incorporating the fast tuning parameter selection procedure, the methods are denoted as Fast-AdapDISCOM and Fast-AdapDISCOM-Huber, respectively.
\subsection{Design}
Following  the multimodal framework proposed by \citet{yu_optimal_2020}, we assume that $\textbf{X}$ is a block-missing matrix comprising three modalities. Specifically, $\textbf{X}$ consists of $p=300$ predictors, evenly distributed across the three modalities ($100$ predictors per modality). The training sample size is $n,$ with $n/4$ missing values in each modality. That is, the training dataset
is composed of $n/4$ samples with complete observations, $n/4$ samples with observations from the first and the second modalities, $n/4$ samples with observations from the first and the third modalities, and $n/4$ samples with missing observations only from the first modality. See the supplementary material for an illustration of the missingness pattern. We consider different sample sizes $n \in \lbrace 40, 120, 200, 280, 360, 440, 520 \rbrace,$ allowing us to evaluate model performance in both high-dimensional $(n < p)$ and standard  $(n \geq p)$ settings. The validation and test sets consist of $200$ and $400$ samples with complete observations, respectively. The coefficient vector is $\boldsymbol{\beta}^0,$ with each modality containing five nonzero predictors.
\begin{equation*}
    \boldsymbol{\beta}^0 = (\underbrace{0.5, 0.5, 0.5, 0.5, 0.5}_{5}, \underbrace{0, \ldots, 0}_{95}, \underbrace{0.5, 0.5, 0.5, 0.5, 0.5}_{5}, \underbrace{0, \ldots, 0}_{95}, \underbrace{0.5, 0.5, 0.5, 0.5, 0.5}_{5}, \underbrace{0, \ldots, 0}_{95}).
\end{equation*}
We consider six distinct block-missing scenarios to analyze the effect of multimodal heterogeneous data structure and heterogeneous measurement error contamination across modalities. In \textbf{Scenario I}, the modalities are homogeneous and subject to the same level of measurement error. Specifically, the three block-missing modalities are generated by the same multivariate Gaussian distribution $\mathcal{N}(0,\Sigma = (\sigma_{jt}^2)_{1 \leq j,t \leq p}),$ where $\sigma_{jt}^2 = 0.6^{|j-t|}.$ The additive measurement error is identical across modalities and takes values $\tau^2 \in \lbrace 0, 0.2, 0.4, 0.6, 0.8 \rbrace.$ This scenario is designed to isolate the impact of missing data and measurement error in the absence of structural heterogeneity across modalities. Note that when $\tau^2 = 0,$ the data are only subject to missingness, corresponding to the design considered by \citet{yu_optimal_2020}. 

In \textbf{Scenario II}, the three modalities exhibit structural heterogeneity, while the measurement errors remain homogeneous across modalities, following the same pattern as in \textbf{Scenario I}. Specifically, each modality is generated from a distinct multivariate Gaussian distribution with a different correlation structure.
\begin{itemize}
    \item Modality~1 follows the same correlation structure as in \textbf{Scenario I}, with $\Sigma_{jt} = 0.6^{|j-t|}.$
    \item Modality~2 features a block-diagonal correlation structure, constructed by repeating $p/5$ times the $5 \times 5$ block matrix $\Sigma_{\text{bloc}}= 0.15 \cdot \mathbf{1}_{5 \times 5} + 0.85 \cdot \mathbb{I}_5,$
    as in Example 2 of the simulation design in \citet{yu_optimal_2020}
    \item  Modality~3, inspired by \citet{Scom2023}, is generated from a covariance matrix defined as the Kronecker product of two matrices: $\Sigma = \Sigma_1 \otimes \Sigma_2$, where $\Sigma_1$ is $3 \times 3$ matrix with entries $\Sigma_{1jt} = 0.8^{|j-t|},$ and $\Sigma_2$ is a larger $p/3 \times p/3$ matrix with entries $\Sigma_{2jt} = 0.3^{|j-t|}.$  
\end{itemize}
This scenario is designed to capture the impact of structural heterogeneity across modalities while keeping the measurement errors consistent. In \textbf{Scenario III}, the data modalities are generated with distinct correlation structures, as in \textbf{Scenario II}, but are further affected by heterogeneous measurement errors. Specifically, the three block-missing modalities retain the same correlation structures described in \textbf{Scenario II}, while the additive measurement errors differ across modalities. For the first and second modalities, the measurement error variances are fixed at $\tau^2=0.2$ and $\tau^2=0.5,$ respectively, whereas for the third modality, the measurement error variance varies over $\tau^2 \in \lbrace 0, 0.2, 0.4, 0.6, 0.8 \rbrace.$  In \textbf{Scenario IV}, the data are affected only by measurement errors, with no missing values, and the modalities are generated following the correlation structures described in \textbf{Scenario II}. In all four scenarios, the random error $\boldsymbol{\varepsilon}$ follows a standard Gaussian distribution. In \textbf{Scenario V}, the first two modalities are simulated from a Gaussian mixture distribution $(\rho \cdot \mathcal{N} (\boldsymbol{0}, \mathbf{I}_{200}) + (1- \rho) \cdot \mathcal{N} (\boldsymbol{0}, 0.5\cdot\mathbf{I}_{200}), \text{ where } \rho = 0.03 ) $, while the third modality follows a multivariate Student distribution with 5 degrees of freedom ($t_5 (\boldsymbol{0}, 0.6\Sigma),$ with entries $\Sigma_{ij} = 0.6^{|i-j|}).$ \textbf{Scenario VI} retains the same multimodal structure as \textbf{Scenario V} but assumes the model error follows a Student distribution with 5 degrees of freedom. These two scenarios are designed to assess the performance of the methods when both the predictors and the model error exhibit heavy-tailed distributions.  We also considered an additional scenario to assess the sensitivity of the AdapDISCOM family of methods. In particular, we included a scenario, \textbf{Scenario VII}, where the data are affected solely by missing values, with the proportion of complete cases varying to induce different levels of missingness. The full description of the design and the corresponding results are provided in the supplementary materials.

Each simulation is repeated $B=100$ times, and variability in the results is visualized using boxplots. We compare the performance of AdapDISCOM-based methods against nine competing approaches: the DISCOM-based methods (DISCOM, Fast-DISCOM, DISCOM-Huber, and Fast-DISCOM-Huber), SCOM \citep{Scom2023}, CocoLasso \citep{datta_cocolasso_2017}, two LASSO-based imputation methods (LASSO-mean and LASSO-SVD), and the complete-case LASSO \citep{lasso_tib}, which uses only complete observed samples. Performance is evaluated using four metrics:
\begin{itemize}
    \item Mean Squared Error (MSE), defined as $\text{MSE}=n^{-1}\| \boldsymbol{y}_{Test} - \widehat{\boldsymbol{y}}_{Test}\|_2^2,$ which assesses the average prediction error magnitude, where $\boldsymbol{y}$ is the response and $\widehat{\boldsymbol{y}}$ its prediction.
    \item Coefficient of determination $(\text{R}^2),$ given by $\text{R}^2 = 1 - \| \boldsymbol{y}_{Test} - \widehat{\boldsymbol{y}}_{Test}\|_2^2/ \|\boldsymbol{y}_{Test} - \overline{\boldsymbol{y}}_{Test}\|_2^2$ which measures the proportion of variance explained by the predictions.
    \item Bias, computed as $\| \widehat{\boldsymbol{\beta}}-\boldsymbol{\beta}^0\|_2,$ which quantifies the estimation error between the true parameter vector and its estimate \citep{hastie2009elements}.
    \item F1-score, defined as $F_1 = \frac{2\text{TP}}{2\text{TP} + \text{FP} + \text{FN}},$ which evaluates the variable selection accuracy, where $\text{TP}, \ FP, \ \text{and} \ FN$ denote the numbers of true positives, false positives, and false negatives, respectively \citep{taha2015metrics}.
\end{itemize}

\subsection{Results}
\begin{center}
\begin{figure}[hbt!]
\includegraphics[width=\linewidth]{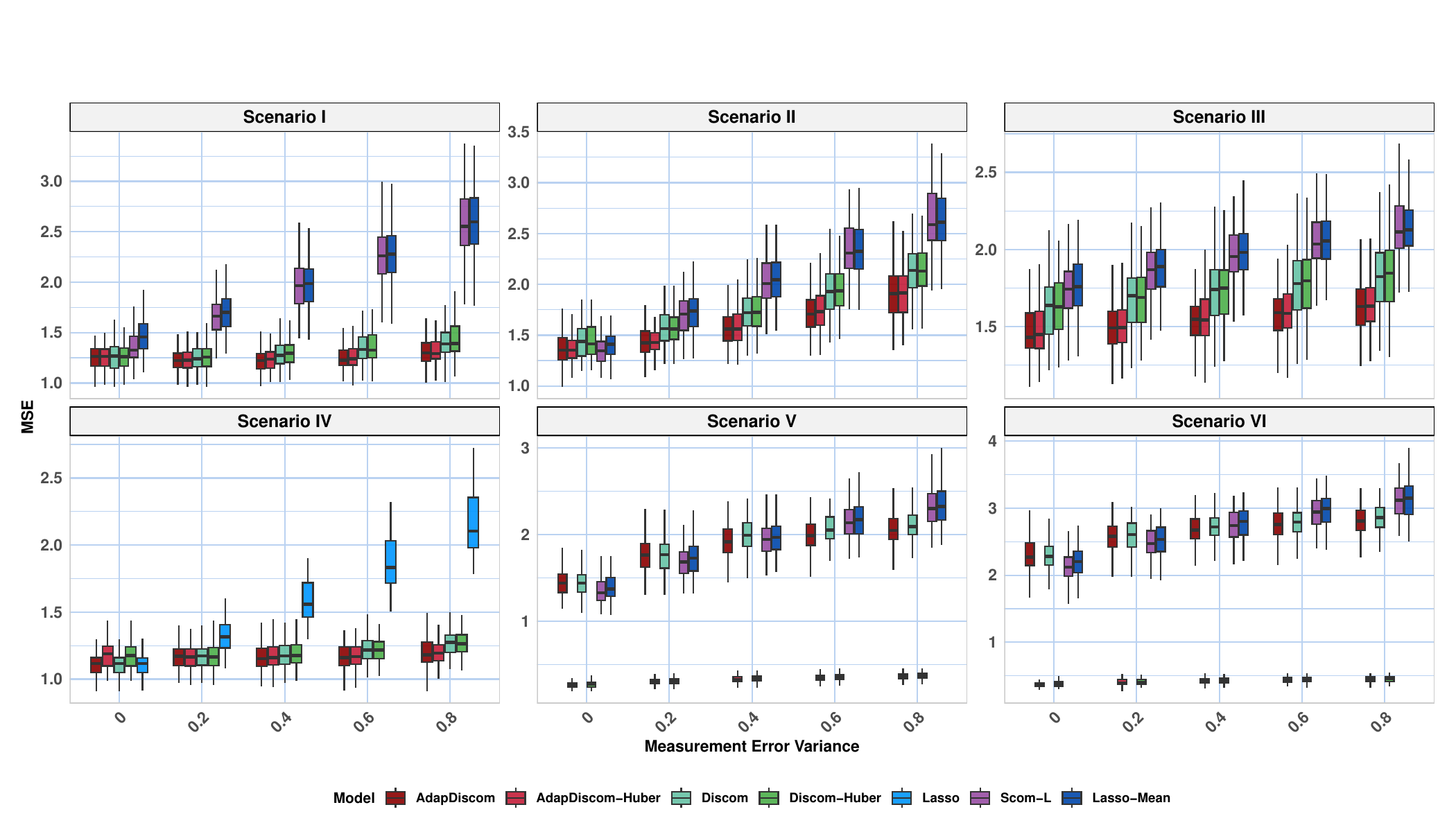}
    \caption{Mean squared error (MSE) of the different methods across the first six scenarios and varying levels of measurement error variance with $n=400.$ In \textbf{Scenario IV}, which involves only measurement error, imputation-based methods are excluded and SCOM is equivalent to LASSO. Results for LASSO and other baseline methods are provided in the supplementary material to improve the readability and highlight the performance of our proposed methods.} \label{fig:mse_sim}
\end{figure}
\end{center}

\begin{center}
\begin{figure}[hbt!]
\includegraphics[width=\linewidth]{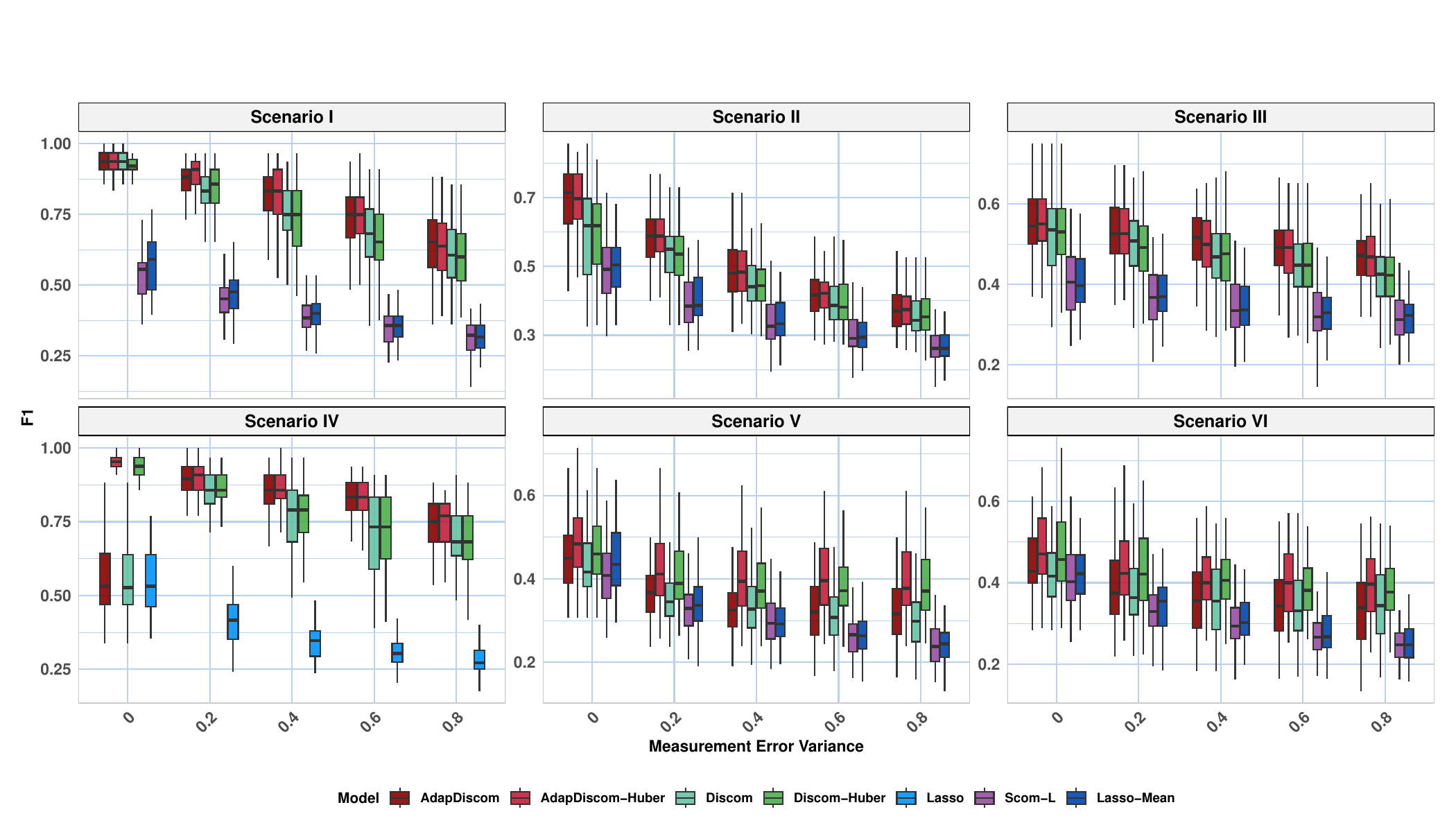}
    \caption{F1-score of the different methods across the first six scenarios and varying levels of measurement error variance with $n=400.$ \textbf{Scenario IV}, which involves only measurement error, imputation-based methods are excluded and SCOM is equivalent to LASSO. Results for LASSO and other baseline methods are provided in the supplementary material to improve the readability and highlight the performance of our proposed methods.} \label{fig:F1_sim}
\end{figure}
\end{center}

We conducted extensive simulations to evaluate the performance of our proposed AdapDISCOM method under a variety of settings combining missing data, measurement error, heterogeneous modality structures, and heavy-tailed distributions. To highlight the key findings, \textbf{Figures \ref{fig:mse_sim}} and \textbf{Figures \ref{fig:F1_sim}} report the MSE and F1-score results for a subset of methods and metrics at $n=440.$ Due to space constraints and to emphasize the superior performance of AdapDISCOM, we omitted certain baseline methods from the main manuscript. In particular, the complete-case LASSO was excluded owing to its poor performance, which obscured differences among competitive approaches. Similarly, among the imputation-based methods, only one representative was retained, as their results were largely similar. Results for the Fast variants of DISCOM and AdapDISCOM, as well as analyses across varying sample sizes, are presented in the supplementary material, where consistent conclusions are observed.

Across all scenarios, we observe that both MSE and bias increase with the magnitude of the additive measurement error and decrease as the sample size grows. These findings underscore the severe impact that missing data and measurement error can have on standard estimators and illustrate the limitations of conventional methods, such as LASSO, in handling these pervasive challenges. In contrast, AdapDISCOM effectively mitigates these detrimental effects, ensuring more reliable inference even in challenging settings.

Regarding MSE, AdapDISCOM consistently achieves the lowest prediction error across all scenarios, regardless of whether contamination is absent, present in the form of missing data, measurement error, or both. In \textbf{Scenario I} (homogeneous modalities without structural heterogeneity), AdapDISCOM performs comparably to DISCOM when no measurement error is present. However, as measurement error increases, AdapDISCOM clearly outperforms DISCOM, with the performance gap widening as contamination grows. In \textbf{Scenario II} (heterogeneous modality structures with homogeneous measurement error), AdapDISCOM maintains a clear advantage over all methods, including DISCOM and SCOM, and the margin of superiority grows with increasing error. In \textbf{Scenario III} (heterogeneous structures and heterogeneous measurement error across modalities, the most realistic setting), the benefits of AdapDISCOM are even more pronounced. This scenario highlights the importance of modality-specific weighting, which allows AdapDISCOM to leverage less-contaminated modalities more effectively. \textbf{Scenario IV}, which involves only additive measurement error without missing data, further illustrates AdapDISCOM’s strength in correcting for measurement error, outperforming both LASSO and CocoLasso (results for these baselines are detailed in the supplementary material). Finally, in the presence of heavy-tailed distributions (\textbf{Scenario V} and \textbf{Scenario VI}), AdapDISCOM — particularly the Huber variants — demonstrates superior robustness compared to all competing methods.

In terms of F1-score, AdapDISCOM performs comparably to DISCOM in the absence of measurement error and homogeneous structures, both outperforming the other methods. However, as soon as measurement error is introduced, AdapDISCOM consistently achieves better variable selection accuracy across all scenarios.

Regarding bias, the superiority of AdapDISCOM is not always systematic. In scenarios without measurement error, SCOM occasionally matches or slightly outperforms AdapDISCOM, likely due to SCOM’s focus on mitigating missingness and methods like CocoLasso being more directly tailored to bias correction. Nevertheless, AdapDISCOM achieves a better overall trade-off between prediction accuracy and bias, making it a robust and versatile choice across diverse settings.

AdapDISCOM can become computationally intensive when the number of hyperparameters, which scales with the number of modalities, is large. Fortunately, the Fast-AdapDISCOM variant alleviates this limitation while maintaining comparable performance. Moreover, thanks to its adaptive strategy, which adjusts the lower bound of the hyperparameter based on the minimal sample size of each modality, Fast-AdapDISCOM can even outperform Fast-DISCOM in terms of computational efficiency when its lower bound exceeds that of Fast-DISCOM (see Supplementary figure).

In summary, our simulation study demonstrates the robustness and versatility of AdapDISCOM, which effectively mitigates the combined effects of missing data and measurement error across both simple and complex, realistic scenarios. AdapDISCOM consistently outperforms not only methods specialized for missing data (DISCOM, SCOM) but also those designed specifically for measurement error (CocoLasso), even in scenarios without contamination. These results underscore the practical value of AdapDISCOM in real-world applications where such data imperfections are prevalent.

\section{Application to Multimodal High-Dimensional Data from the ADNI Cohort}
Alzheimer’s disease represents a major global public health challenge. In 2018, Alzheimer’s Disease International estimated that approximately 50 million individuals worldwide were living with dementia \citep{gustavsson2023global}, a number projected to triple by 2050 \citep{prince2015world}. Advancing research on Alzheimer’s disease is thus critical to mitigating its impact on population health and quality of life. In this context, the Alzheimer’s Disease Neuroimaging Initiative (ADNI) \citep{mueller2005alzheimer} has played a pivotal role.

The data used in this section were obtained from the ADNI database (\url{adni.loni.usc.edu}). The ADNI was launched in 2003 as a public-private partnership, led by Principal Investigator Michael W. Weiner, MD. The primary goal of ADNI has been to test whether serial magnetic resonance imaging (MRI), positron emission tomography (PET), other
biological markers, and clinical and neuropsychological assessment can be combined to measure the progression of mild cognitive impairment (MCI) and early Alzheimer’s disease (AD). ADNI is inherently multimodal, comprising structural and functional brain imaging (MRI and PET), genetic data, cerebrospinal fluid (CSF) biomarkers, and neuropsychological measures that can be integrated to assess disease progression. These modalities, however, come from different acquisition techniques and exhibit block-wise missingness (see missing pattern figure in supplementary) and heterogeneous measurement errors, attributable to the high cost of certain tests (e.g., PET scans), variability in data quality, and patient reluctance to undergo invasive procedures (e.g., CSF collection). Developing robust methods capable of addressing these challenges is essential for reliable inference and prediction.

We applied the AdapDISCOM family of methods to evaluate the association between multimodal ADNI data and Mini-Mental State Examination (MMSE) scores, while accounting for the complex structure of missingness and measurement error. MMSE is a widely used 30-point questionnaire designed to assess cognitive function, particularly in older adults or patients suspected of dementia \citep{folstein1975mini}. Specifically, we analyzed data from four modalities: MRI-derived volumetric measures, Flortaucipir PET SUVRs, CSF biomarkers, and genetic variants (SNPs).

After preprocessing and quality control of each modality (details in supplementary), the final dataset comprised 1,211 subjects and 1,748 features: 5 CSF biomarkers, 36 subcortical MRI volumes, 74 PET SUVRs, and 1,633 SNPs (filtered by GWAS $p\text{-value}< 5 \times 10^{-5}).$ The CSF biomarkers included $A\beta 42,$ total tau (T-tau),  phosphorylated tau at threonine 181 (P-tau),  and the ratios T-tau/A$\beta$ 42, and P-tau/A$\beta$ 42. We considered two analysis scenarios:
\begin{itemize}
    \item \textbf{Scenario I}: three modalities (CSF + MRI + PET), analogous to the setup in \citet{yu_optimal_2020}, excluding genetic data.
    \item \textbf{Scenario II}: all four modalities (CSF + MRI + PET + SNPs), to assess the potential confounding effect of genetic variants on MMSE prediction.
\end{itemize}
In \textbf{Scenario I}, 196 subjects had complete data (split into 116 for testing, 40 for tuning, and the remaining 40 plus 1,015 incomplete cases for training), with 53.6\% missingness overall. In \textbf{Scenario II}, only 86 subjects had complete data (split into 40 for testing, 36 for tuning, and 10 plus 1,125 incomplete cases for training), with 53.44\% missingness. For each scenario, hyperparameters were tuned using a 30-point grid based on KKT conditions to define $\lambda_{\max},$ with $\lambda_{\min}$ set as a fraction $(10^{-2})$ of $\lambda_{\max}$ following the GLMnet approach \citep{friedman2010regularization}. Each method was run 40 times per scenario, with performance assessed via MSE and $R^2$ on the test set.

\subsection{Results}
\begin{center}
\begin{figure}[hbt!]
\includegraphics[width=\linewidth]{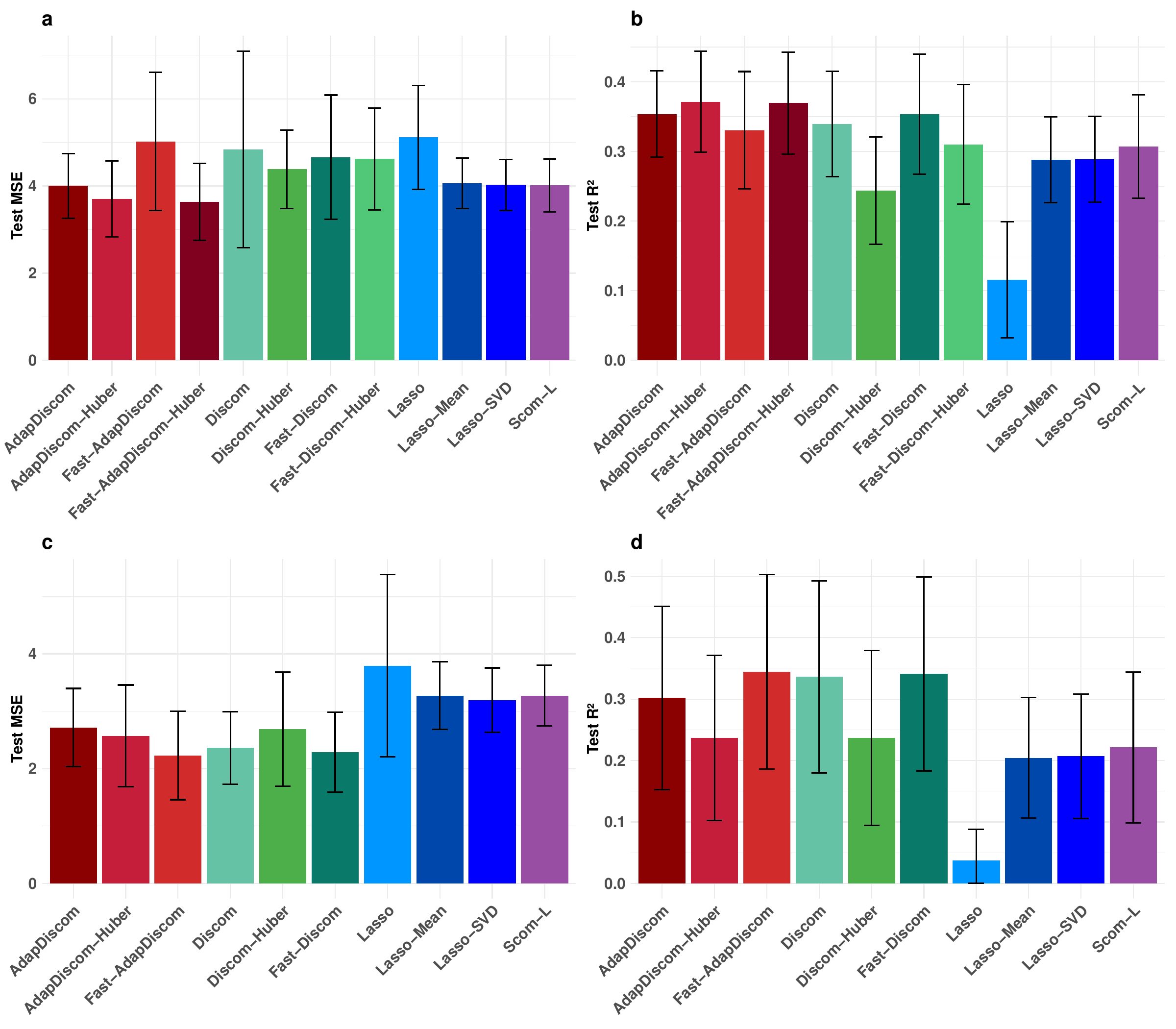}
    \caption{Mean squared error (MSE) and $R^2$ of the different methods on the test set, presented as bar plots with standard deviation error bars. Results for \textbf{Scenario I} (CSF + MRI + PET) are displayed in the top row, and results for \textbf{Scenario II} (CSF + MRI + PET + SNPs) in the bottom row.} \label{fig:real_mse_r2}
\end{figure}
\end{center}

Figure \ref{fig:real_mse_r2} presents the test-set MSE and $R^2$ of all methods. Across both scenarios, AdapDISCOM-based methods outperformed competing approaches. In \textbf{Scenario I}, AdapDISCOM-Huber achieved the best overall performance, while in \textbf{Scenario II}, Fast-AdapDISCOM yielded the lowest MSE and highest $R^2.$ DISCOM-based methods also performed well, particularly in \textbf{Scenario II}, outperforming some AdapDISCOM variants. In contrast, while LASSO complete case and LASSO imputed methods showed acceptable MSE, their $R^2$ values were very low, reflecting their poor predictive utility in this high-dimensional, incomplete-data setting. SCOM performed comparably to imputation-based methods but lagged behind AdapDISCOM and DISCOM, likely due to its lack of covariance regularization and its conceptual similarity to imputation-based approaches. 

\begin{center}
\begin{figure}[hbt!]
\includegraphics[width = 1.1\linewidth]{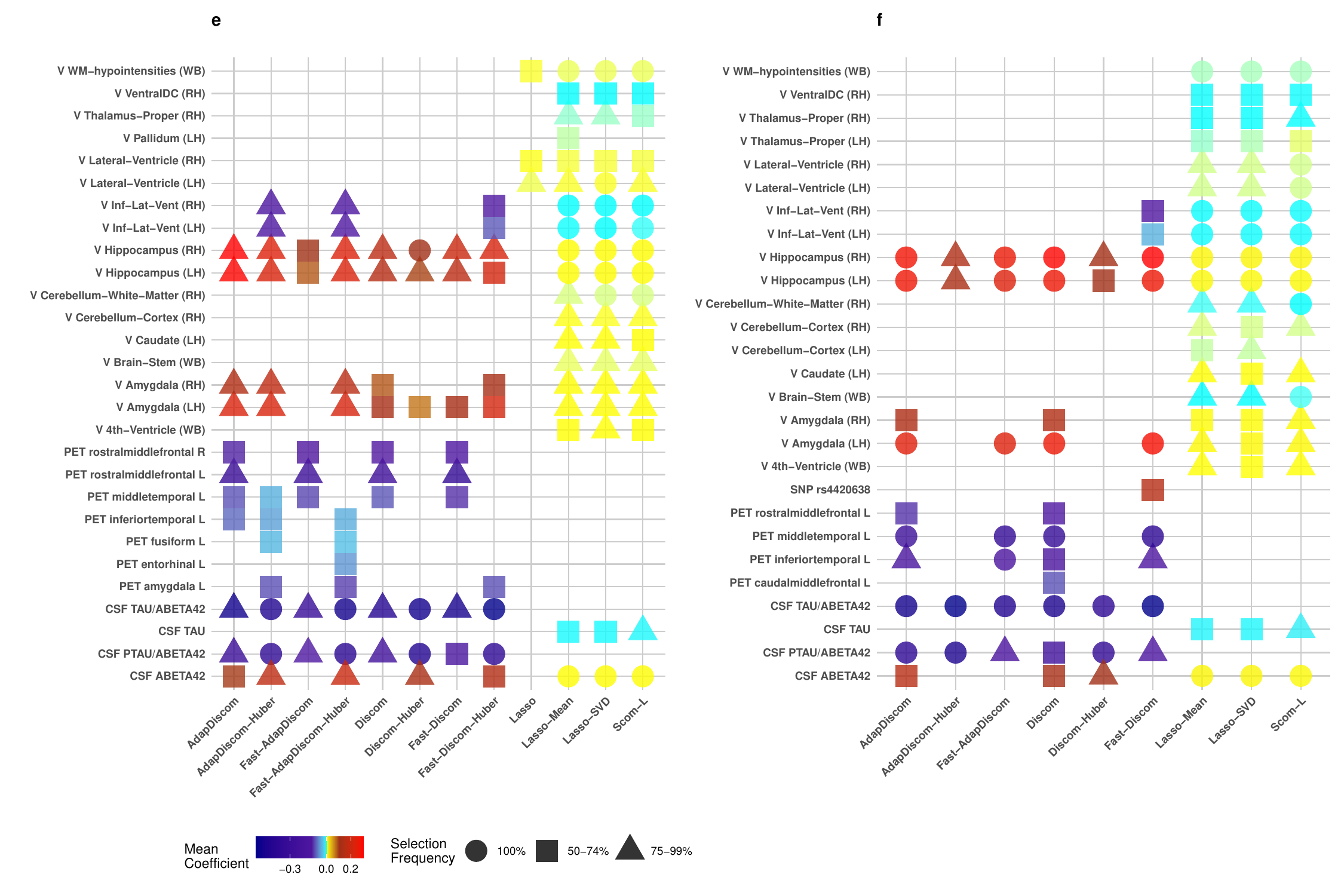}
    \caption{Average effect sizes (colour gradient) of features selected at least 50\% (square), 75\% (triangle), or 100\% (circle) of the time by a given method.} \label{fig:real_beta}
\end{figure}
\end{center}

Figure \ref{fig:real_beta} illustrates variables selected consistently $(\geq 50\%, \geq 75\%, \ \text{and}\ \geq 100\%$ of the time) by each method, along with their effect sizes (color gradient). Results were consistent across scenarios, with slightly more robustly selected variables in \textbf{Scenario I}. We observed two distinct groups of methods: AdapDISCOM/DISCOM and LASSO-based/SCOM. The variables selected by AdapDISCOM and DISCOM differed from those selected by LASSO-based methods and SCOM, which tended to select more variables with negligible or null effects. In contrast, variables selected consistently by AdapDISCOM and DISCOM had substantial, directionally consistent effects across methods, further demonstrating the inefficacy of conventional approaches in the presence of measurement error and missing data.

Both AdapDISCOM and DISCOM consistently identified well-established Alzheimer’s disease biomarkers and brain regions, including  CSF ratios T-tau/A$\beta$ 42 and P-tau/A$\beta$ 42, bilateral hippocampal and amygdalar volumes from MRI, and PET SUVRs in frontal and temporal ROIs (e.g., rostral middle frontal, middle temporal, inferior temporal, and amygdala regions). These findings align with the literature, where elevated CSF ratios are strong indicators of cognitive decline \citep{mckenna2025p, fagan2007cerebrospinal}, and reduced hippocampal and amygdalar volumes are associated with dementia risk \citep{bocchetta2021looking, watson2016subcortical, peng2015correlation}. Likewise, higher Flortaucipir SUVRs in frontal and temporal lobes are associated with lower MMSE performance, indicating a correlation between increased tau deposits and poorer cognitive performance \citep{tsai201918f}. Notably, none of the PET SUVR features were selected by the LASSO-based or SCOM methods, in contrast to AdapDISCOM and DISCOM. In \textbf{Scenario II}, SNPs were selected by our methods, though not systematically across runs.

In summary, AdapDISCOM demonstrated robust predictive performance and reliable variable selection, consistent with established Alzheimer’s biomarkers. Nonetheless, given the limited number of complete cases, these findings should be interpreted cautiously. The primary aim of this analysis was to illustrate the potential of AdapDISCOM rather than to conduct an exhaustive clinical study of Alzheimer’s disease. These results underscore AdapDISCOM’s promise as a practical, effective method for high-dimensional, multimodal data analysis under real-world data imperfections.

\section{Discussion}
We have introduced AdapDISCOM, a novel prediction and variable selection method designed to simultaneously address block-wise missing data and additive measurement error in multimodal settings. Like DISCOM, AdapDISCOM follows a two-step procedure: the first step estimates the predictor covariance matrix, while the second step solves a LASSO-type optimization problem for prediction and selection. However, AdapDISCOM improves upon DISCOM by adaptively weighting the intra-modality covariance matrices, decomposing them into a weighted sum of modality-specific estimates. This weighting explicitly accounts for differences in data-generating mechanisms and the heterogeneity and variability in measurement error magnitudes across modalities. As a result, the estimated covariance matrix remains positive semi-definite and more accurate than the empirical sample covariance. The estimated matrix is then incorporated into the LASSO objective function for joint prediction and variable selection.

Our simulation results clearly demonstrate the ability of AdapDISCOM to mitigate the effects of block-wise missing data, additive measurement error, and their combination. AdapDISCOM consistently outperformed competing methods in scenarios with heterogeneous modality structures and heterogeneous measurement errors, highlighting the importance of explicitly modeling inter-modality heterogeneity and error magnitude variability. These findings underscore the robustness of AdapDISCOM in practical, realistic settings where modalities may contribute unevenly due to differential contamination or data quality.

We further evaluated AdapDISCOM on the task of predicting MMSE scores from multimodal ADNI data, demonstrating its practical utility as a prediction and selection tool in the presence of both block-wise missingness and measurement error. Despite having only a few dozen fully observed samples in the training set, AdapDISCOM achieved superior performance over standard LASSO (both complete-case and imputation-based) and SCOM, producing more robust and accurate predictions. AdapDISCOM methods achieved the lowest MSE and highest $R^2,$ while robustly selecting the same nonzero predictors (e.g., CSF, Amygdala, Hippocampus) with effect directions consistent with the literature. In contrast, imputation-based LASSO and SCOM tended to select more predictors, often with near-zero effects, reducing interpretability and reliability.

While AdapDISCOM shows strong performance, it requires tuning more hyperparameters as the number of modalities increases, which can be computationally expensive in high dimensions. To address this, we proposed Fast-AdapDISCOM, which reduces the number of hyperparameters to the same level as Fast-DISCOM, regardless of the number of modalities, while remaining more computationally efficient. Unlike DISCOM, Fast-AdapDISCOM exploits the adaptive weights specific to each modality and their complete-sample sizes, leading to improved efficiency and scalability.

We also noted that the performance of AdapDISCOM can degrade when predictors and outcomes come from heavy-tailed distributions. To overcome this limitation, we developed AdapDISCOM-Huber, a robust variant that performs particularly well under such complex data conditions. Both simulation and real data analyses confirmed that AdapDISCOM-Huber outperforms other AdapDISCOM variants and standard methods when the data exhibit heavy tails. Additionally, a Fast-AdapDISCOM-Huber version is available, offering a practical combination of robustness and computational efficiency. We recommend Fast-AdapDISCOM-Huber as a robust and scalable option that outperforms existing standard approaches.

All methods in the AdapDISCOM family, including DISCOM, have been implemented and are freely available on \href{https://github.com/AODiakite/AdapDiscom}{GitHub}, and currently under review for inclusion on CRAN.

Several avenues remain open for further extending and generalizing AdapDISCOM. One natural direction is to adapt the method to generalized linear models and longitudinal data, and to explore alternative penalty functions beyond the standard LASSO. Since AdapDISCOM was originally developed under the missing completely at random (MCAR) assumption, another critical area of future research is its extension to settings with missing at random (MAR) or missing not at random (MNAR) mechanisms. Beyond measurement error and missing data, AdapDISCOM could also be enhanced to detect outliers, inspired by the approach in \citep{Barry18082022}. Finally, incorporating expectiles and M-quantiles \citep{barry2023alternative} offers a promising direction, enabling inference beyond the mean and capturing more nuanced features of the response distribution.

\section{Conclusion}
In conclusion, AdapDISCOM offers a robust and flexible framework for prediction and variable selection in the presence of block-wise missing data and measurement error, common challenges in multimodal biomedical studies. Through its adaptive weighting mechanism and robust extensions, AdapDISCOM effectively leverages the heterogeneity across modalities while maintaining computational scalability. Our simulations and real-world application to ADNI data demonstrate that AdapDISCOM, particularly its robust and fast variants, consistently outperforms existing methods in terms of prediction accuracy, variable selection reliability, and interpretability. Future work will focus on extending AdapDISCOM to generalized linear models, alternative loss functions such as expectiles and M-quantiles, and incorporating outlier detection capabilities. These directions highlight the potential of AdapDISCOM as a versatile tool for modern high-dimensional and imperfect data analyses.

\section{Software}
All methods in the AdapDISCOM family, including DISCOM, have been implemented and are freely available on \href{https://github.com/AODiakite/AdapDiscom}{GitHub}, and currently under review for inclusion on CRAN.

\section{Acknowledgements}
This work is supported by the Natural Sciences and Engineering Research Council of Canada through an individual discovery research grant to Amadou Barry. 

This research was enabled in part by support provided by Calcul Québec (\url{https://www.calculquebec.ca/}) and the Digital Research Alliance of Canada (\url{https://alliancecan.ca/en}).

Data collection and sharing for this project was funded by the Alzheimer's Disease Neuroimaging Initiative (ADNI) (National Institutes of Health Grant U01 AG024904) and DOD ADNI (Department of Defense award number W81XWH-12-2-0012). ADNI is funded by the National Institute on Aging, the National Institute of Biomedical Imaging and Bioengineering, and through generous contributions from the following: AbbVie, Alzheimer’s Association; Alzheimer’s Drug Discovery Foundation; Araclon Biotech; BioClinica, Inc.; Biogen; Bristol-Myers Squibb Company; CereSpir, Inc.; Cogstate; Eisai Inc.; Elan Pharmaceuticals, Inc.; Eli Lilly and Company; EuroImmun; F. Hoffmann-La Roche Ltd and its affiliated company Genentech, Inc.; Fujirebio; GE Healthcare; IXICO Ltd.; Janssen Alzheimer Immunotherapy Research \& Development, LLC.; Johnson \& Johnson Pharmaceutical Research \& Development LLC.; Lumosity; Lundbeck; Merck \& Co., Inc.; Meso Scale Diagnostics, LLC.; NeuroRx Research; Neurotrack Technologies; Novartis Pharmaceuticals Corporation; Pfizer Inc.; Piramal Imaging; Servier; Takeda Pharmaceutical Company; and Transition Therapeutics. The Canadian Institutes of Health Research is providing funds to support ADNI clinical sites
in Canada. Private sector contributions are facilitated by the Foundation for the National Institutes of Health (\url{www.fnih.org}). The grantee organization is the Northern California Institute for Research and Education, and the study is coordinated by the Alzheimer’s Therapeutic Research Institute at the University of Southern California. ADNI data are disseminated by the Laboratory for Neuro Imaging at the University of Southern California.

\clearpage
\section{Appendix}
In this appendix, we first provide the proofs of the theorems stated in the main manuscript. The second section presents simulation results that were not included in the main text. Finally, the last section details the preprocessing and quality control procedures applied to the real data.

\subsection{Proof}
In this section, we present the proofs of the results stated in the manuscript, which are a generalization of the DISCOM method. With close attention, it can be seen that, with very few exceptions, the proofs of the generalization are identical to those presented by  \citet{yu_optimal_2020}. We add them here for completeness. 
\begin{proof}[\textbf{Proof of Proposition 1}] 
We want to optimize the following problem
\begin{equation*}
\begin{split}
    & \min_{\alpha_k,\alpha_C, \alpha_p}\E\Big[\|\widehat{\Sigma} - \Sigma\|^2_F\Big]  \\
    & \text{s.t.} \ \widehat{\Sigma} = \sum_{k=1}^K \alpha_k \widetilde{\Sigma}_{I_k} + \alpha_C \widetilde{\Sigma}_{C} +  \sum_{k=1}^K (1-\alpha_k) \gamma \mathbb{I}_p.
\end{split}
\end{equation*} 
Noting that $\Sigma = \sum_{k=1}^K \Sigma_{I_k} + \Sigma_{C}, \; \E[\widetilde{\Sigma}_{I_k}] = \Sigma_{I_k}$ and $\E[\widetilde{\Sigma}_C] = \Sigma_{C}$ we have
\begin{align*}
{} & \E [\|\widehat{\Sigma} - \Sigma\|^2_F] = \E \Bigg[\Big\|\sum_{k=1}^K \alpha_k \widetilde{\Sigma}_{I_k} + \alpha_C \widetilde{\Sigma}_{C} +  \sum_{k=1}^K (1-\alpha_k) \gamma \mathbb{I}_p - \Sigma \Big\|_F^2\Bigg]\\ 
 & =  \E \Bigg[ \Big\| \sum_{k=1}^K \alpha_k (\widetilde{\Sigma}_{I_k} - \Sigma_{I_k}) + \sum_{k=1}^K (1-\alpha_k) (\gamma\mathbb{I}_p -\Sigma_{I_k}) \Big\|^2_F\Bigg] + \E [\| \alpha_C\widetilde{\Sigma}_C - \Sigma_C \|^2_F] \\
 & = \sum_{k=1}^K  \alpha_k^2 \E[\|(\widetilde{\Sigma}_{I_k} - \Sigma_{I_k})\|^2_F] + \sum_{k=1}^K (1-\alpha_k)^2 \|(\gamma\mathbb{I}_p - \Sigma_{I_k})\|^2_F + \E[\|(\alpha_C\widetilde{\Sigma}_C-\Sigma_C)\|^2_F.
\end{align*}
The last expression is obtained by applying the independence property. The expression is more easily seen by taking $K = 3.$

By minimizing the expression $\sum_{k=1}^K (1-\alpha_k)^2 \|(\gamma\mathbb{I}_p - \Sigma_{I_k})\|^2_F$ with respect to $gamma$ we get the optimal value: $\gamma* = \frac{1}{p} \sum_{k=1}^{K} \frac{(1- \alpha_k^*)^2}{\sum_{t=1}^{K}(1-\alpha_t^*)^2} \Tr(\widetilde{\Sigma}_{I_k}).$ The optimal value of $\alpha_C$ is obtained by minimizing $\E[\|(\alpha_C\widetilde{\Sigma}_C-\Sigma_C)\|^2_F]$ and then we have $\alpha_C^* = \frac{\|\Sigma_{C}\|^2_F}{\|\Sigma_{C}\|^2_F + \delta_{C}^2}.$ Replacing $\gamma$ by its optimal value in the objective function and taking the derivative with respect to $\alpha_k,$ we obtain, for $k\in \lbrace 1,2,\ldots, K\rbrace,$ the optimal value $\alpha_k^*  = \frac{\theta_{I_k}^2}{\theta_{I_k}^2 + \delta_{I_k}^2}.$ Thus, the optimal value of $\alpha_p$ is $\alpha_p^* = \gamma^* \sum_{k=1}^K (1 - \alpha_k^*) = \gamma^* \sum_{k=1}^K\frac{\delta_{I_k}^2}{\theta_{I_k}^2 + \delta_{I_k}^2}.$

At the optimum, the value of the objective function is equal to $\sum_{k=1}^K\frac{\theta_{I_k}^2}{\theta_{I_k}^2 + \delta_{I_k}^2} \delta_{I_k}^2 + \frac{\|\Sigma_{C}\|^2_F}{\delta_{C}^2 + \|\Sigma_{C}\|^2_F} \delta_{C}^2$ which is less than $\sum_{k=1}^K \delta_{I_k}^2 + \delta_{C}^2.$ Since $\E[\|\widetilde{\Sigma} - \Sigma\|^2_F] = \sum_{k=1}^K \delta_{I_k}^2 + \delta_{C}^2,$
we have $\E[\|\widetilde{\Sigma}^* - \Sigma\|^2_F] \leq \E[\|\widetilde{\Sigma} - \Sigma\|^2_F].$
\end{proof}

In the following, to simplify the notation, we omit the exponent $k$ in the expressions such as the variance and the sample size. From now on, variance $\sigma_{jt}^k$ will be denoted by $\sigma_{jt},$ and $n_{jt}^k = n_{jt}, \, n_j^k = n_j.$
\begin{proof}[\textbf{Proof of Theorem 1}] 
The proof of this theorem is based on the following lemma:
\begin{lemma}\label{lem_ravi} \citet{Ravikumar_tail_bound2011}
Consider a zero-mean random vector $(X_1, \ldots, X_p)$ with covariance $\Sigma$ such that each $X_j/\sqrt{\sigma_{jj}}$ is sub-Gaussian with parameter L. Given n i.i.d. samples, the associated sample covariance $\widehat{\Sigma}$ satisfies the following tail bound
\begin{equation*}
    P\Big(|\widehat{\sigma}_{jt} - \sigma_{jt}| \ge \delta \Big)  \leq 4 \exp\Big\lbrace - \frac{n\delta^2}{128(1+L^2)^2\max_j(\sigma_{jj})^2} \Big\rbrace,
\end{equation*}
for all $\delta \in (0, 8\max_j(\sigma_{jj})(1+4L^2)).$
\end{lemma}
\textbf{Convergence rate of} \; $\widetilde{\Sigma}.$ \;
To begin, we assume that $X_j$ is standardized that is $\sigma_{jj} =1.$ Now, under condition $(\textbf{A1}),$ and $\delta = \nu_1\sqrt{\frac{\log p}{n_{jt}}}$ and $\nu_1 = 8\sqrt{6}(1+4L^2) \max_j(\sigma_{jj}^k) = 8\sqrt{6}(1+4L^2)$ such that condition of Lemma \ref{lem_ravi} is verified. Note that the expression of $\delta$ is obtained when $\min_{j,t}n_{jt} > 6\log p.$  Then:
\begin{equation*} \label{eq1}
\begin{split}
P(|\widetilde{\sigma}_{jt} - \sigma_{jt}| \ge \delta) & \leq 4 \exp\Big\lbrace - \frac{n_{jt}\delta^2} { 128(1+4L^2)^2\max_j(\sigma_{jj})^2}\Big\rbrace \\
 & \leq 4 \exp\Big\lbrace - \frac{n_{jt} 384 (1+4L^2)^2 \max_j(\sigma_{jj})^2 \frac{\log p}{n_{jt}}} { 128(1+4L^2)^2\max_j(\sigma_{jj})^2}\Big\rbrace \\
 & \leq 4 \exp\Big\lbrace - \frac{384 (1+4L^2)^2 \max_j(\sigma_{jj})^2 } { 128(1+4L^2)^2\max_j(\sigma_{jj})^2} \log p \Big\rbrace \\
 & \leq 4 p^{-3} \quad \quad \text{ for any} \; j,t \in \lbrace 1, 2, \ldots, p \rbrace.
\end{split}
\end{equation*}
Hence, with $\nu_2 = 4,$ and the condition $\min_{j,t}n_{jt} > 6\log p,$ we have:
\begin{equation} \label{first_result}
\max_{j,t} P\Big(|\widetilde{\sigma}_{jt} - \sigma_{jt}| \ge \nu_1\sqrt{\frac{\log p}{n_{jt}}}\Big)  \leq   \frac{\nu_2}{p^3}.
\end{equation}
Now using result of \ref{first_result} and the inequality $P(\| A \|_{\max} \ge \delta) \leq \sum_{i,j} P(|a_{ij}| \ge \delta)$ we have
\begin{equation*} 
 P\Big(\|\widetilde{\Sigma} - \Sigma \|_{\max} \ge \nu_1\sqrt{\frac{\log p}{\min_{j,t}n_{jt} }}\Big)  \leq   \frac{\nu_2}{p^3} \cdot p^2 = \frac{\nu_2}{p}.
\end{equation*}
\textbf{Convergence rate of} \; $\widetilde{\text{C}}.$ \; The random variable $y/\sqrt{\Var(y)}$ is sub-Gaussian with parameter $\frac{L}{\min\lbrace 1, \sqrt{\Var(y)} \rbrace}$ and accordingly the upper bound of $\delta$ in Lemma \ref{lem_ravi} is $8\max\lbrace 1, \Var(y)\rbrace \frac{L^2}{\min\lbrace 1, \sqrt{\Var(y)} \rbrace}.$ Now taking $\delta = \nu_3\sqrt{\frac{\log p}{n_{j}}}, \, \nu_3 = 16(1 + 4\frac{L^2}{\min\lbrace 1, \Var(y)\rbrace}) \max\lbrace 1, \Var(y)\rbrace, \; \nu_4 = 4$ and obsserving that $\min_j n_j \geq min_{j,t}n_{jt} > 6 log p > 4 log p$, then applying Lemma \ref{lem_ravi} gives as:
\begin{align*} 
\max_{j} P\Big(|\widetilde{\text{c}}_{j} - \text{c}_{j}| \ge \nu_3\sqrt{\frac{\log p}{n_{j}}}\Big) & \leq   \frac{\nu_4}{p^2} \\ 
 P\Big(\|\widetilde{\text{C}} - \text{C} \|_{\max} \ge \nu_3\sqrt{\frac{\log p}{\min_{j}n_{j} }}\Big) & \leq   \frac{\nu_4}{p^2} \cdot p = \frac{\nu_4}{p}.
\end{align*}

\textbf{Convergence rate of}  \; $\widetilde{\boldsymbol{\beta}}.$ \; Given that $\widetilde{\boldsymbol{\beta}}$ is the solution to  
\begin{equation*}
\argmin_{\boldsymbol{\beta}\in \mathbb{R}^p} \frac{1}{2} \boldsymbol{\beta}\transpose \Bigg[ \sum_{k=1}^K \alpha_k \widetilde{\Sigma}_{I_k} + \alpha_C\widetilde{\Sigma}_{C} +   \sum_{k=1}^K (1-\alpha_k) \times \sum_{k=1}^K \frac{(1-\alpha_k)^2}{\sum_{t=1}^K (1-\alpha_t)^2} \frac{\Tr(\widetilde{\Sigma}_{I_k})}{p}  \mathbb{I}_p \Bigg] \boldsymbol{\beta} -\widetilde{\text{C}}\transpose\boldsymbol{\beta}+ \lambda\|\boldsymbol{\beta}\|_1,
\end{equation*}
we have
\begin{equation*}
    \frac{1}{2}\widetilde{\boldsymbol{\beta}}\transpose\widehat{\Sigma}\widetilde{\boldsymbol{\beta}} - \widetilde{\text{C}}\transpose \widetilde{\boldsymbol{\beta}} + \lambda\|\widetilde{\boldsymbol{\beta}}\|_1 \leq 
     \frac{1}{2}\boldsymbol{\beta}^0\transpose\widehat{\Sigma}\boldsymbol{\beta}^0 - \widetilde{\text{C}}\transpose \boldsymbol{\beta}^0 + \lambda\|\boldsymbol{\beta}^0\|_1.
\end{equation*}
Now with further expansion, and taking into account the KKT condition, $\| \widehat{\Sigma}\widetilde{\boldsymbol{\beta}} -\widetilde{\text{C}} \|_{\max} \leq \lambda,$  we have: 
\begin{align*} 
& 2\lambda \|\widetilde{\boldsymbol{\beta}}\|_1 \leq \boldsymbol{\beta}^0\transpose\widehat{\Sigma}\boldsymbol{\beta}^0 - \widetilde{\boldsymbol{\beta}}\transpose\widehat{\Sigma}\widetilde{\boldsymbol{\beta}} + 2 \widetilde{\text{C}}\transpose (\widetilde{\boldsymbol{\beta}}- \boldsymbol{\beta}^0) + 2\lambda\|\boldsymbol{\beta}^0\|_1\\
& = (\boldsymbol{\beta}^0-\widetilde{\boldsymbol{\beta}})\transpose\widehat{\Sigma}\boldsymbol{\beta}^0 + \widetilde{\boldsymbol{\beta}}\transpose\widehat{\Sigma}(\boldsymbol{\beta}^0-\widetilde{\boldsymbol{\beta}}) + 2 \widetilde{\text{C}}\transpose (\widetilde{\boldsymbol{\beta}}- \boldsymbol{\beta}^0) + 2\lambda\|\boldsymbol{\beta}^0\|_1 \\
& = (2 \widetilde{\text{C}} - \widehat{\Sigma}\widetilde{\boldsymbol{\beta}} - \widehat{\Sigma}\boldsymbol{\beta}^0)\transpose (\widetilde{\boldsymbol{\beta}}- \boldsymbol{\beta}^0) + 2\lambda\|\boldsymbol{\beta}^0\|_1 \\
& \leq (\|\widetilde{\text{C}}  - \widehat{\Sigma}\widetilde{\boldsymbol{\beta}}\|_{\max} + \|\widetilde{\text{C}}  - \widehat{\Sigma}\boldsymbol{\beta}^0\|_{\max}) \cdot \|\boldsymbol{\beta}^0-\widetilde{\boldsymbol{\beta}}\|_1 + 2\lambda\|\boldsymbol{\beta}^0\|_1 \\
& \leq (\lambda + \|\widetilde{\text{C}}  - \widehat{\Sigma}\boldsymbol{\beta}^0\|_{\max}) \cdot \|\boldsymbol{\beta}^0-\widetilde{\boldsymbol{\beta}}\|_1 + 2\lambda\|\boldsymbol{\beta}^0\|_1.
\end{align*}
As assumed in the theorem, if the tuning parameter $\lambda = 2\|\widetilde{\text{C}}  - \widehat{\Sigma}\boldsymbol{\beta}^0\|_{\max},$ we have
\begin{equation*}
    2\lambda \|\widetilde{\boldsymbol{\beta}}\|_1 \leq 3/2\cdot\lambda\|\boldsymbol{\beta}^0-\widetilde{\boldsymbol{\beta}}\|_1 + 2 \lambda \|\boldsymbol{\beta}^0\|_1
\end{equation*}
Hence,
\begin{equation*}
    \|\widetilde{\boldsymbol{\beta}} - \boldsymbol{\beta}^0\|_1 \leq 4(\|\widetilde{\boldsymbol{\beta}} - \boldsymbol{\beta}^0\|_1 + \|\boldsymbol{\beta}^0\|_1 - \|\widetilde{\boldsymbol{\beta}}\|_1).
\end{equation*}
Let $\delta = \widetilde{\boldsymbol{\beta}} - \boldsymbol{\beta}^0.$ Since for $j\in J^c, \ |\widetilde{\boldsymbol{\beta}}_j - \boldsymbol{\beta}^0_j| + |\boldsymbol{\beta}^0_j| - |\widetilde{\boldsymbol{\beta}}_j| = 0$ it holds that
\begin{equation}\label{trans_delta}
    \|\widetilde{\boldsymbol{\beta}} - \boldsymbol{\beta}^0\|_1 = \|\delta_J\|_1 +  \|\delta_{J^c}\|_1 \leq 4(\|\delta_J\|_1 + \|\beta_J^0\|_1 - \|\widetilde{\beta}_J\|_1) \leq 8 \|\delta_J\|_1.
\end{equation}
Hence, $\|\delta_{J^c}\|_1 \leq 7  \|\delta_J\|_1.$ 

Now consider, $\widehat{\Sigma} = \sum_{k=1}^K \alpha_k \widetilde{\Sigma}_{I_k} + \alpha_C\widetilde{\Sigma}_{C} +   \sum_{k=1}^K (1-\alpha_k) \mathbb{I}_p,$ where $1-\alpha_k = \bigO(\sqrt{\log p/\min_j n_j}), \; k \in\lbrace 1,2, \ldots K \rbrace$ and $1-\alpha_C = \bigO(\sqrt{\log p/\min_{j,t}n_{jt}}).$ Denote events $\mathcal{A}=\lbrace \|\widehat{\Sigma} - \Sigma \|_{\max} \ge \nu_1\prime\sqrt{\log p/\min_{j,t}n_{jt} }\rbrace$ and $\mathcal{B} =\lbrace \|\widetilde{\text{C}} - \text{C} \|_{\max} \ge \nu_3\sqrt{\log p/\min_{j}n_{j} } \rbrace .$ 
Based on the definition of $\widehat{\Sigma} = (\sigma_{jt})_{j,t=1}^p,$ we know that
\begin{equation*}
  \widehat{\sigma}_{jt} - \sigma_{jt} =
    \begin{cases}
      \sum_{k \ne k'}^K (1-\alpha_k)  & \text{if  $j=t$ and $j \in $ modality $k'$;} \\
      \alpha_k \widetilde{\sigma}_{jt} - \sigma_{jt} & \text{if  $j \ne t$ \ ($j$ and $t$ are in the same modality);}\\
      \alpha_C \widetilde{\sigma}_{jt} - \sigma_{jt} & \text{if  $j \ne t$ \ ($j$ and $t$ are in different modalities).}\\
    \end{cases}       
\end{equation*}
Thus, if $j \ne t$ and the predictors $j$ and $t$ are in the same modality, with probability at least $1-\nu_2/p^3,$ we have:
\begin{equation*} 
\begin{split}
\lvert \widehat{\sigma}_{jt} - \sigma_{jt} \rvert = \lvert \alpha_k \widetilde{\sigma}_{jt} - \sigma_{jt} \rvert & \leq \alpha_k \lvert \widetilde{\sigma}_{jt} - \sigma_{jt} \rvert + (1-\alpha_k)\lvert \sigma_{jt} \rvert \\
& \leq \alpha_k \lvert \widetilde{\sigma}_{jt} - \sigma_{jt} \rvert + (1-\alpha_k) \\
& \leq \alpha_k \nu_1 \sqrt{\log p/\min_j n_j}+ (1-\alpha_k) \quad \text{derived from above results} \\
& \leq  \nu_1 \sqrt{\log p/\min_j n_j}+ (1-\alpha_k).  \\
\end{split}
\end{equation*}
Similarly, if $j\ne t$ and the predictors $j$ and $t$ are in different modalities, with probability at least $1 -\nu_2/p^3,$ we have
\begin{equation*} 
\lvert \widehat{\sigma}_{jt} - \sigma_{jt} \rvert = \lvert \alpha_C \widetilde{\sigma}_{jt} - \sigma_{jt} \rvert \leq \nu_1 \sqrt{\log p/\min_{j,t} n_{jt}}+ 1-\alpha_C.
\end{equation*}
Therefore, there exists two constants $\nu_1^\prime$ and $\nu_2$ such that
\begin{equation*} 
 P\Big(\|\widehat{\Sigma} - \Sigma \|_{\max} \ge \nu_1\prime\sqrt{\log p/\min_{j,t}n_{jt} }\Big)  \leq  \nu_2/p.
\end{equation*}
From previous results and the above convergence rate of $\|\widehat{\Sigma} - \Sigma \|_{\max},$ we have $P(\mathcal{A}\cap \mathcal{B})\ge 1 - (\nu_2 + \nu_4)/p.$ In events $\mathcal{A}$  and $\mathcal{B},$ we have
\begin{equation*} \label{eq11}
\begin{split}
\|\widetilde{\text{C}} - \widehat{\Sigma}\boldsymbol{\beta}^0\|_{\max}  & \leq \|\widetilde{\text{C}} - \text{C} \|_{\max} +
\|\widehat{\Sigma}- \Sigma \|_{\max} \|\boldsymbol{\beta}^0 \|_1 \\
& \leq (\nu_3 + \nu_1\prime \|\boldsymbol{\beta}^0 \|_1) \sqrt{\log p/\min_{j,t}n_{jt}}.
\end{split}
\end{equation*}
Therefore, we have $\|\widetilde{\text{C}} - \widehat{\Sigma}\boldsymbol{\beta}^0\|_{\max} = \bigO_p\Big(\|\boldsymbol{\beta}^0 \|_1\sqrt{\log p/\min_{j,t}n_{jt}}\Big).$ 
Furthermore, under the condition $(\textbf{A2}),$  if the event $\mathcal{A}$ occurs, we have
\begin{equation*}
    \frac{\delta\transpose\widehat{\Sigma}\delta}{\delta\transpose\delta} = \frac{\delta\transpose\Sigma\delta}{\delta\transpose\delta} + \frac{\delta\transpose(\widehat{\Sigma}-\Sigma)\delta}{\delta\transpose\delta} \ge m-64s \|\widehat{\Sigma}-\Sigma\|_{\max} \ge m-64s\nu_1\prime \sqrt{\log p/\min_{j,t}n_{jt}}.
\end{equation*}
If we assume that $s\nu_1\prime \sqrt{\log p/\min_{j,t}n_{jt}} = \smallO(1)$ or $\min_{j,t}n_{jt} > (128\nu_1\prime/m)^2(s^2\log p,)$ we have
\begin{equation}\label{eqS1}
    \frac{\delta\transpose\widehat{\Sigma}\delta}{\delta\transpose\delta} \ge m - m/2 = m/2 > 0,
\end{equation}
for sufficiently large $s, \ p,$  and $\min_{j,t}n_{jt}.$ On the other hand, we have
\begin{equation} \label{eqS2}
\begin{split}
\delta\transpose\widehat{\Sigma}\delta & \leq  \|\widehat{\Sigma}(\widetilde{\boldsymbol{\beta}} - \boldsymbol{\beta}^0)\|_{\max}\|\delta\|_1 \leq 
(\|\widehat{\Sigma}\widetilde{\boldsymbol{\beta}} - \widetilde{C}\|_{\max} + \|\widetilde{C} - \widehat{\Sigma}\boldsymbol{\beta}^0) \|_{\max}) \|\delta\|_1 \\
& \leq (\lambda + \|\widetilde{C} - \widehat{\Sigma}\boldsymbol{\beta}^0 \|_{\max}) \|\delta\|_1 = 1.5\lambda\|\delta\|_1
\end{split}
\end{equation}
Therefore, by (\ref{eqS1}) and (\ref{eqS2}), we have $\frac{m}{2}\|\delta\|_2^2 \leq \delta\transpose\widehat{\Sigma}\delta \leq 1.5\lambda\|\delta\|_1 \leq 12 \lambda\|\delta_J\|_1 \leq 12\lambda \sqrt{s}\|\delta\|_2.$ Hence, $\|\delta\|_2\leq 24\lambda\sqrt{s}/m.$ Therefore,
$\|\widetilde{\boldsymbol{\beta}} - \boldsymbol{\beta}^0\|_2 = \bigO_p(\sqrt{s}\lambda)=\bigO_p(\|\boldsymbol{\beta}^0\|_1\sqrt{s\log p /\min_{j,t}n_{jt}}).$ This completes the proof.

\end{proof}
\begin{proof}[\textbf{Proof of Theorem 2}] 
\textbf{Convergence rate of} \; $\breve{\Sigma}.$ \; We can prove this Theorem under the conditions (under condition $(\textbf{A3})$ and $\min_{j,t,k}n_{jt}^k \geq 24 \log p$) stated in the main manuscript and by relying on Theorem 5 in \citet{heavy_tail_fan2016}. For all $j,t, \in \lbrace 1,2, \ldots, p \rbrace,$ we know that
\begin{equation*}
    P \Big( |\breve{\sigma}_{jt} - \sigma_{jt}| \geq \text{Q}_1 \sqrt{\frac{\log p}{n_{jt}}}\Big) \leq \frac{2}{p^3}. \; \text{Then,} \; 
    \max_{j,t} P \Big( |\breve{\sigma}_{jt} - \sigma_{jt}| \geq \text{Q}_1 \sqrt{\frac{\log p}{n_{jt}}}\Big) \leq \frac{2}{p^3}.
\end{equation*}
Considering the fact that $\min_{j,t}n_{jt} < n_{jt} \forall j, t, \in \lbrace 1,2, \ldots, p \rbrace $ we have
\begin{equation*}
    P \Big( \| \breve{\Sigma} - \Sigma \|_{\max} \geq \text{Q}_1 \sqrt{\frac{\log p}{\min_{j,t} n_{jt}}} \Big) = 
    P \Big( \max_{j,t} |\breve{\sigma}_{jt} - \sigma_{jt}| \geq \text{Q}_1 \sqrt{\frac{\log p}{\min_{j,t} n_{jt}}} \Big) \leq \frac{2}{p}.
\end{equation*}
\textbf{Convergence rate of} \; $\breve{\text{C}}.$ Noting that: for each $j \in \lbrace 1,2, \ldots, p \rbrace$ and $i\in S_j,$ we have $\Var(x_{ij}y_i)\leq \E(x_{ij}^2y_i^2) = \sqrt{\E(x_{ij}^4) \E(y_i^4)}\leq 2 (Q_1 + Q_2)^2.$ Now noting, $\text{H}_j = (\text{Q}_1 + \text{Q}_2) \sqrt{n_j/\log p}$ for each $j \in \lbrace 1,2, \ldots, p,$ and using Theorem 5 in \citet{heavy_tail_fan2016}, we have
\begin{equation*}
    \max_{j} P \Big( |\breve{c}_{j} - c_{j}| \geq 8 (\text{Q}_1 + \text{Q}_2) \sqrt{\frac{\log p}{n_{j}}}\Big) \leq \frac{2}{p^2}.
\end{equation*}
Again, since $\min_{j}n_{j} < n_{j} \forall j \in \lbrace 1,2, \ldots, p \rbrace $ we have
\begin{equation*}
 P \Bigg( \| \breve{\text{C}} - \text{C} \|_{\max} \geq 8 (\text{Q}_1 + \text{Q}_2)  \sqrt{\frac{\log p}{\min_{j} n_{j}}} \Bigg) \leq \frac{2}{p}. 
\end{equation*}
This completes the proof.
\end{proof}
\begin{proof}[\textbf{Proof of Theorem 3}] 
\textbf{Sub-Gaussian Case}.  \; By the KKT condition, we know that $\widetilde{\boldsymbol{\beta}}$ is a solution to the optimization problem if and only if there exists a subgradient $\gamma\in\mathbb{R}^p$ such that
\begin{equation*}
    \widetilde{C} - \widehat{\Sigma}\widetilde{\boldsymbol{\beta}} = \lambda\gamma,
\end{equation*}
 where for each $j \in \lbrace 1, 2, \ldots, p \rbrace, \ \gamma_j = \sign(\widetilde{\beta}_j)$ if $\widetilde{\beta}_j\ne 0,$ and $\gamma_j \in [-1, 1]$ if $\widetilde{\beta}_j=0.$

 We can construct a point $\widetilde{\boldsymbol{\beta}}\in\mathbb{R}^p$ by letting $\widetilde{\boldsymbol{\beta}}_J = (\widehat{\Sigma}_{JJ})^{-1}\widetilde{C}_J-\lambda(\widehat{\Sigma}_{JJ})^{-1}\cdot\sign(\boldsymbol{\beta}^0)$ and $\boldsymbol{\beta}^0_{J^c} = 0.$ Define events $\mathcal{A}_1 = \lbrace  \|\widetilde{\boldsymbol{\beta}}_J - \boldsymbol{\beta}_J^0\|_{\max} < \beta_{\min}^0\rbrace$ and 
 $\mathcal{A}_2 = \lbrace  \|\widetilde{C}_{J^c} - \widehat{\Sigma}_{J^cJ}\widetilde{\boldsymbol{\beta}}_J\|_{\max}\leq \lambda\rbrace.$ If events $\mathcal{A}_1$ and $\mathcal{A}_2$ hold, we can check that $\widetilde{\boldsymbol{\beta}}$ is a solution and $\sign(\widehat{\boldsymbol{\beta}}) =\sign(\boldsymbol{\beta}^0).$ To prove the theorem, we only need to show that $P(\mathcal{A}_1)\longrightarrow 1$ and $P(\mathcal{A}_2)\longrightarrow 1,$ as $\min_{j,t}n_{jt} \longrightarrow \infty$ and $p \longrightarrow \infty$

\textbf{Step 1}: show the upper bound of $\|(\widehat{\Sigma}_{JJ})^{-1}\|_{\infty}.$ 

Denote $V =\|(\widehat{\Sigma}_{JJ})^{-1}\|_{\infty}.$ Since
\begin{align*} 
&  \|(\widehat{\Sigma}_{JJ})^{-1} - (\Sigma)_{JJ}^{-1}\|_{\infty} \leq  \|(\Sigma)_{JJ}^{-1}\|_{\infty} \cdot \|(\widehat{\Sigma}_{JJ})^{-1}\|_{\infty} \cdot \|( \widehat{\Sigma}_{JJ} - \Sigma_{JJ})\|_{\infty}\\ 
& \leq \|(\Sigma)_{JJ}^{-1}\|_{\infty} \cdot ( \|(\Sigma)_{JJ}^{-1}\|_{\infty} + \|(\widehat{\Sigma}_{JJ})^{-1} - (\Sigma)_{JJ}^{-1}\|_{\infty}) \cdot\|( \widehat{\Sigma}_{JJ} - \Sigma_{JJ})\|_{\infty} \\
& = V (V + \|(\widehat{\Sigma}_{JJ})^{-1} - (\Sigma)_{JJ}^{-1}\|_{\infty}) \cdot \| \widehat{\Sigma}_{JJ} - \Sigma_{JJ}\|_{\infty},
\end{align*}
we have,
\begin{equation*}
    \|(\widehat{\Sigma}_{JJ})^{-1} - (\Sigma)_{JJ}^{-1}\|_{\infty} \leq \frac{V^2\| \widehat{\Sigma}_{JJ} - \Sigma_{JJ}\|_{\infty} }{1-V\| \widehat{\Sigma}_{JJ} - \Sigma_{JJ}\|_{\infty}} \leq \frac{sV^2\| \widehat{\Sigma}_{JJ} - \Sigma_{JJ}\|_{\max} }{1-sV\| \widehat{\Sigma}_{JJ} - \Sigma_{JJ}\|_{\max}} 
\end{equation*}
and
\begin{equation*}
    \|(\widehat{\Sigma}_{JJ})^{-1}\|_{\infty} \leq V + \frac{sV^2\| \widehat{\Sigma}_{JJ} - \Sigma_{JJ}\|_{\max} }{1-sV\| \widehat{\Sigma}_{JJ} - \Sigma_{JJ}\|_{\max}} = \frac{V}{1-sV \| \widehat{\Sigma}_{JJ} - \Sigma_{JJ}\|_{\max}}.
\end{equation*}
\textbf{Step 2}: show the upper bound of $\| \widetilde{C}_{J} - \widehat{\Sigma}_{JJ}\boldsymbol{\beta}^0_J\|_{\max}.$ We have
\begin{equation*} 
\begin{split}
\| \widetilde{C}_{J} - \widehat{\Sigma}_{JJ}\boldsymbol{\beta}^0_J\|_{\max} & \leq \| \widetilde{C}_{J} - C_{J}\|_{\max} +  \| (\Sigma_{JJ} - \widehat{\Sigma}_{JJ})\boldsymbol{\beta}^0_J\|_{\max}\\
 & \leq \| \widetilde{C}_{J} - C_{J}\|_{\max} +  \| (\Sigma_{JJ} - \widehat{\Sigma}_{JJ})\|_{\infty}\|\boldsymbol{\beta}^0_J\|_{\max}\\
 & \leq \| \widetilde{C}_{J} - C_{J}\|_{\max} + s \beta_{\max}^0\| (\Sigma_{JJ} - \widehat{\Sigma}_{JJ})\|_{\infty}.
\end{split}
\end{equation*}
\textbf{Step 3}: show that $P(\mathcal{A}_1\longrightarrow 1)$ as $\min_{j,t}n_{jt}$ and $p \longrightarrow \infty.$

Define event $\mathcal{A}_3=\lbrace \|\widehat{\Sigma} - \Sigma \|_{\max} \leq \nu_1^\prime\sqrt{\log p /\min_{j,t}n_{jt}} \rbrace$ and 
$\mathcal{A}_4=\lbrace \|\widetilde{C} - C \|_{\max} \leq \nu_3\sqrt{\log p /\min_{j}n_{j}}\rbrace.$ By Theorem 1, we know that $P(\mathcal{A}_3)\longrightarrow 1$ and $P(\mathcal{A}_4)\longrightarrow 1$ as $p\longrightarrow\infty.$ If events $\mathcal{A}_3$ and $\mathcal{A}_4$ occur, since $\frac{1+s\beta_{\max}^0}{\lambda}\sqrt{\frac{\log p}{\min_{j,t}n_{jt}}}\longrightarrow 0,$ if $sV\sqrt{\log p/\min_{j,t}n_{jt}}\longrightarrow 0$ or the following condition holds
\begin{equation*}
    \|(\widehat{\Sigma}_{JJ})^{-1}\|_{\infty} \cdot \sqrt{\frac{s^2\log p}{\min_{j,t}n_{jt}}} \leq \frac{\eta}{\nu_1\prime (4+\eta)},
\end{equation*}
we have
\begin{align*} 
&  \|\widetilde{\boldsymbol{\beta}}_J - \boldsymbol{\beta}_J^0\|_{\max} =  \| (\widehat{\Sigma}_{JJ})^{-1}\widetilde{C}_{J} - \lambda(\widehat{\Sigma}_{JJ})^{-1}\cdot  \sign(\boldsymbol{\beta}_J^0) - \boldsymbol{\beta}_J^0 \|_{\max}\\ 
& \leq \| (\widehat{\Sigma}_{JJ})^{-1}\widetilde{C}_{J}  - \boldsymbol{\beta}_J^0 \|_{\max} + \lambda \|(\widehat{\Sigma}_{JJ})^{-1}\|_{\infty} \\
& \leq (\|\widetilde{C}_{J} -\widehat{\Sigma}_{JJ}\boldsymbol{\beta}_J^0\|_{\max} + \lambda) \cdot \|(\widehat{\Sigma}_{JJ})^{-1}\|_{\infty} \\
& \leq (\|\widetilde{C}_{J} - C_J\|_{\max} + s\boldsymbol{\beta}_{\max}^0\|\widehat{\Sigma}_{JJ} - \Sigma_{JJ} \|_{\max} + \lambda) \cdot \frac{V}{1 - sV\|\widehat{\Sigma}_{JJ} - \Sigma_{JJ} \|_{\max}} \\
& \leq \frac{2\lambda V}{1 - sV\|\widehat{\Sigma}_{JJ} - \Sigma_{JJ} \|_{\max}} \leq 4 \lambda V,\\ 
\end{align*}
for sufficiently large $p$ and $\min_{j,t}n_{jt}.$ Therefore, if $\lambda V/\boldsymbol{\beta}^0_{\min}\longrightarrow 0,$ we have $P(\mathcal{A}_1) = 1 - P (\lbrace\|\widetilde{\boldsymbol{\beta}}_J - \boldsymbol{\beta}_J^0\|_{\max} \ge \boldsymbol{\beta}^0_{\min}\rbrace) \longrightarrow 1.$

\textbf{Step 4}: We have
\begin{align*} 
&  \| \widehat{\Sigma}_{J^c J}(\widehat{\Sigma}_{JJ})^{-1} - \Sigma_{J^c J}(\Sigma_{JJ})^{-1} \|_{\infty} \\ 
& \leq \| \Sigma_{J^c J} ((\widehat{\Sigma}_{JJ})^{-1} - (\Sigma_{JJ})^{-1}) \|_{\infty} +  \| (\widehat{\Sigma}_{J^c J} - \Sigma_{J^c J}) (\widehat{\Sigma}_{JJ})^{-1} \|_{\infty} \\ 
& \leq \| \Sigma_{J^c J} (\Sigma_{JJ})^{-1}\|_{\infty} \cdot \|\Sigma_{JJ} - \widehat{\Sigma}_{JJ}\|_{\infty} \cdot \|  (\widehat{\Sigma}_{JJ})^{-1} \|_{\infty} \\ 
& \quad\quad\quad + \|  (\widehat{\Sigma}_{JJ})^{-1} \|_{\infty} \cdot \| \widehat{\Sigma}_{J^c J} - \Sigma_{J^c J} \|_{\infty} \\
& \leq \|(\widehat{\Sigma}_{JJ})^{-1} \|_{\infty} \cdot (\| \widehat{\Sigma}_{JJ} - \Sigma_{JJ} \|_{\infty} + \| \widehat{\Sigma}_{J^c J} - \Sigma_{J^c J} \|_{\infty} ) \\
& \leq \frac{2sV \|\widehat{\Sigma} - \Sigma \|_{\max} }{1- sV \|\widehat{\Sigma} - \Sigma\|_{\max} }.
\end{align*}
\textbf{Step 5}: We show that $P(\mathcal{A}_2\longrightarrow 1)$ as $\min_{j,t}n_{jt}$ and $p \longrightarrow \infty.$

Since $\widetilde{\boldsymbol{\beta}}_J = (\widehat{\Sigma}_{JJ})^{-1}\widetilde{C}_J-\lambda(\widehat{\Sigma}_{JJ})^{-1}\cdot\sign(\boldsymbol{\beta}^0),$ we have
\begin{align*} 
&  \| \widetilde{C}_{J^c} - \widehat{\Sigma}_{J^c J}\widetilde{\boldsymbol{\beta}}_J \|_{\max} \leq  \| \widetilde{C}_{J^c} - \widehat{\Sigma}_{J^c J} (\widehat{\Sigma}_{JJ})^{-1}\widetilde{C}_J\|_{\max} + \lambda \|  \widehat{\Sigma}_{J^c J} \widehat{\Sigma}_{JJ})^{-1} \|_{\max} \\
& \leq \| \widetilde{C}_{J^c} - C_{J^c}\|_{\max} +  \|( \Sigma_{J^c J} (\Sigma_{JJ})^{-1} -  \widehat{\Sigma}_{J^c J}(\widehat{\Sigma}_{JJ})^{-1})C_{J^c}\|_{\max} \\
& \quad\quad\quad + \|\widehat{\Sigma}_{J^c J} (\widehat{\Sigma}_{JJ})^{-1} (C_{J} - \widetilde{C}_{J}) \|_{\max} + \lambda  \| \widehat{\Sigma}_{J^c J} (\widehat{\Sigma}_{JJ})^{-1}  \|_{\infty} \\
&  \leq \underbrace{ \| \widetilde{C}_{J^c} - C_{J^c}\|_{\max} }_{(I)} + \underbrace{ s \boldsymbol{\beta}_{\max}^{0}\|\widehat{\Sigma} - \Sigma \|_{\max} \cdot (1 + \|\widehat{\Sigma}_{J^c J} (\widehat{\Sigma}_{JJ})^{-1}\|_{\infty} )   }_{(II)} \\
&  \quad\quad\quad + \underbrace{ (\lambda + \| \widetilde{C}_{J} - C_{J}\|_{\max} \cdot \|\widehat{\Sigma}_{J^c J} (\widehat{\Sigma}_{JJ})^{-1}\|_{\infty} }_{(III)} . 
\end{align*}
Assuming that $\mathcal{A}_3$ and $\mathcal{A}_4$ occur, we know that
\begin{align*} 
  (I) & \leq \nu_3 \sqrt{(\log  p/\min_j n_j)} \leq \nu_3 \sqrt{(\log  p/\min_{j,t} n_{jt}} \\
 (II) &  \leq \nu_1^\prime s \boldsymbol{\beta}_{\max}^{0} \sqrt{(\log  p/\min_{j,t} n_{jt}} \cdot (2 - \eta + \frac{2sV \|\widehat{\Sigma} - \Sigma \|_{\max}}{1 - sV\|\widehat{\Sigma} - \Sigma \|_{\max}  } )\\
 (III) & \leq (\lambda + \nu_3 \sqrt{(\log  p/\min_j n_j)} ) \cdot (1 - \eta + \frac{2sV \|\widehat{\Sigma} - \Sigma \|_{\max}}{1 - sV\|\widehat{\Sigma} - \Sigma \|_{\max}  } ).\\
\end{align*}
Since $\frac{1+ s\boldsymbol{\beta}_{\max}^{0}}{\lambda}\sqrt{\frac{\log  p}{\min_{j,t} n_{jt}}} \longrightarrow 0,$ \; if \; $sV\sqrt{\frac{\log  p}{\min_{j,t} n_{jt}}} \longrightarrow 0$  or we assume that
\begin{equation*}
    \| (\Sigma_{JJ})^{-1}\|_{\infty} \cdot \sqrt{\frac{s^2\log  p}{\min_{j,t} n_{jt}}} \leq \frac{\eta}{ \nu^\prime (4 + \eta)},
\end{equation*}
we have
\begin{equation*}
    \frac{\|\widetilde{C}_{J^c} - \widehat{\Sigma}_{J^c J}\widetilde{\boldsymbol{\beta}}_J \|_{\max}} {\lambda} \leq \frac{(I)}{\lambda} + \frac{(II)}{\lambda} + \frac{(III)}{\lambda}
    \leq \frac{\eta}{4} + \frac{\eta}{4} + 1 - \frac{\eta}{2} = 1,
\end{equation*}
for sufficiently large $p$ and $\min_{j,t} n_{jt}.$

Therefore, $P(\mathcal{A}_2) \longrightarrow 1$ as $\min_{j,t} n_{jt} \longrightarrow \infty$ and $p\longrightarrow \infty.$ This complete the proof.

\textbf{Heavy-Tailed Case}  \; The proof is similar to the \textbf{Sub-Gaussian Case}. We need to define $\mathcal{A}_5 = \lbrace \|\breve{\Sigma} - \Sigma \|_{\max} \leq Q_1^\prime \sqrt{\log p /\min_{j,t}n_{jt}} \rbrace$ and $\mathcal{A}_6=\lbrace \|\breve{C} - C \|_{\max}  \leq 8(Q_1 + Q_2)\sqrt{\log p /\min_{j}n_{j}} \rbrace.$ By Theorem 3, we know that $P(\mathcal{A}_5) \longrightarrow 1$ and $P(\mathcal{A}_6) \longrightarrow 1$ as $p\longrightarrow \infty.$ If events $\mathcal{A}_5$ and $\mathcal{A}_6$ occur, we can show that
$P(\lbrace \|\breve{\boldsymbol{\beta}}_J - \boldsymbol{\beta}^0_J\|_{\max} < \boldsymbol{\beta}^0_{\min}\rbrace) \longrightarrow 1$ and $P(\lbrace \|\breve{C}_{J^c} - \widehat{\Sigma}_{J^c J}\breve{\boldsymbol{\beta}}_J\|_{\max} \leq \lambda \rbrace) \longrightarrow 1$ as $\min_{j,t} n_{jt} \longrightarrow \infty$ and $p\longrightarrow \infty.$

\end{proof}

\subsection{Preprocessing and Quality Control}

\subsubsection{Preprocessing and Quality Control of MRI Data}

Structural T1-weighted (T1w) MRI images from ADNI participants were preprocessed using fMRIPrep v20.2.3, which integrates the FreeSurfer v7.3.2 pipeline for anatomical processing. The workflow included several standard preprocessing and quality control steps, detailed below.

\textbf{Anatomical preprocessing}

The anatomical sub-workflow of fMRIPrep performed the following steps:

\begin{itemize}
    \item \textbf{Brain extraction}  was conducted using the ANTs antsBrainExtraction.sh method \citep{avants2008symmetric, tustison2021antsx}.
    \item \textbf{Tissue segmentation}   of gray matter, white matter, and cerebrospinal fluid was performed using FSL FAST \citep{zhang2002segmentation}.
    \item \textbf{Spatial normalization}   of gray matter, white matter, and cerebrospinal fluid was performed using FSL FAST to the MNI152 template was carried out with ANTs registration algorithms \citep{avants2008symmetric, tustison2021antsx}.
\end{itemize}

A detailed description of this workflow is available at: \url{https://fmriprep.org/en/latest/workflows.html#preprocessing-of-structural-mri}

The resulting brain mask from this preprocessing was then used as input to the FreeSurfer recon-all pipeline.

\textbf{FreeSurfer processing}

The FreeSurfer pipeline implemented the following steps:
\begin{itemize}
    \item \textbf{Bias field correction}, using the N4ITK algorithm \citep{tustison2010n4itk}.
    \item \textbf{Segmentation of subcortical white matter and deep gray matter structures},  according to the Aseg atlas \citep{fischl2002whole, fischl2004sequence}.
    \item \textbf{Tessellation of the gray/white matter boundary}  and \textbf{cortical surface reconstruction} \citep{dale1999cortical, fischl2000measuring}.
\end{itemize}

The final outputs consisted of volumetric segmentations, cortical surface meshes, and derived morphometric measures. Specifically, we extracted:

\begin{itemize}
    \item Regional  \textbf{cortical thickness (CTh)} measurements in each region of the Desikan-Killiany (DK) atlas \citep{desikan2006automated}.
    \item Regional  \textbf{subcortical volumes}, as defined by the Aseg atlas \citep{fischl2002whole}.
\end{itemize}

These measures were then used for downstream analyses.

\subsubsection{Preprocessing and Quality Control of Flortaucipir PET (SUVRs)}

Flortaucipir (AV-1451) PET data from ADNI were processed according to the acquisition and analysis protocols described by the UC Berkeley group (see ADNI methods repository). Below, we summarize the key steps of the preprocessing and quality control (QC) procedures.

\textbf{Data acquisition and initial preprocessing}. 
We downloaded raw flortaucipir PET images in their fully preprocessed format from the ADNI repository (series description: AV1451 Coreg, Avg, Std Img and Vox Siz, Uniform Resolution). These images had already undergone standard ADNI preprocessing, including spatial standardization and initial intensity normalization to a cerebellar cortex region \citep{jagust2015alzheimer}. However, following UC Berkeley recommendations, we replaced this initial normalization with a subject-specific normalization based on FreeSurfer-defined reference regions in native space to reduce warping-induced noise and improve regional specificity.

For each subject, we selected the T1-weighted MRI scan closest in time to the flortaucipir PET acquisition. MRI images were bias-field corrected and segmented using FreeSurfer v7.1.1 to define cortical and subcortical regions of interest (ROIs) in native space.

\textbf{Image coregistration and normalization}. PET images were linearly registered to the corresponding MRI using ANTs and SPM. T1 MRI images were first registered (linear + nonlinear) to the ADNI template using ANTs, and PET images were then aligned to the T1 MRI and warped into template space using the T1-to-template transformation. This ensured consistent spatial normalization across modalities.

After registration, PET images were stripped of non-brain tissue, including meninges, using the brain masks generated during MRI segmentation.

\textbf{SUVR computation}. SUVRs (standardized uptake value ratios) were computed in native space within FreeSurfer-defined ROIs. Specifically, regional flortaucipir uptake values were normalized to the inferior cerebellar grey matter, identified using a combination of SUIT cerebellar atlas regions and FreeSurfer cerebellar grey masks, excluding regions prone to off-target binding \citep{diedrichsen2006spatially}. This reference region is preferred to mitigate dorsal cerebellar contamination and improve comparability across subjects.

For regional quantification, we used ROIs defined by FreeSurfer segmentation, including cortical and subcortical areas, as well as composite regions approximating Braak stages (Braak 1–6) and a meta-temporal composite ROI \citep{maass2017comparison, scholl2016pet}. SUVRs were calculated as the mean uptake within each target ROI divided by the mean uptake in the inferior cerebellar grey reference region.

\textbf{Partial volume correction}. 
To mitigate partial volume effects, we also computed SUVRs corrected using the Geometric Transfer Matrix (GTM) method \citep{rousset1998correction}, modelling both ROIs and off-target binding regions such as the choroid plexus \citep{maass2017comparison}. Both PVC and non-PVC SUVR datasets were generated, with the non-PVC dataset using MRI scans closest in time to each PET scan and the PVC dataset using a baseline MRI across all time points.

\textbf{Partial volume correction}. 
All processed PET images were visually inspected to assess the quality of coregistration to both the individual T1 MRI and the ADNI template. Misalignments, artefacts, and registration failures were flagged and, where necessary, reprocessed.

\subsubsection{Preprocessing and Quality Control of Genomic Data}
Genomic data for this study were derived from the ADNI dataset and generated using the Illumina Omni 2.5M microarray platform, which assays approximately 2.4 million single nucleotide polymorphisms (SNPs) across the genome. The initial dataset included 812 individuals and 2,379,855 SNPs in total.

\textbf{Quality control procedures}. 
We applied a rigorous quality control (QC) pipeline using the software PLINK v1.9. The following filters were sequentially applied:

\begin{itemize}
    \item Individuals with a missing genotype rate exceeding 10\% were excluded (\textendash mind 0.1).
    \item SNPs with a call rate lower than 95\% were removed (\textendash geno 0.05).
    \item SNPs with a minor allele frequency (MAF) below 1\% were excluded to eliminate rare variants likely to contribute noise (\textendash maf 0.01).
    \item SNPs deviating from Hardy-Weinberg equilibrium (HWE) at a significance level of $p < 1 \times 10^{-6}$ were filtered out (\textendash hwe 1e-6).
\end{itemize}

\textbf{Linkage disequilibrium pruning}. 
To reduce redundancy among highly correlated SNPs due to linkage disequilibrium (LD), we conducted LD pruning with the following parameters: a sliding window of 50 kilobases, a step size of 5 SNPs, and an $r^2$ threshold of 0.2. This process removed SNPs in strong LD and retained a subset of approximately independent markers for downstream analyses. After these QC and pruning steps, a total of 265,303 high-quality SNPs were retained, corresponding to an overall genotyping rate of 99.8\%.

\textbf{Population structure assessment}. 
To account for potential population stratification, we performed principal component analysis (PCA) on the pruned dataset. The first 10 principal components were extracted and later used as covariates in association analyses to adjust for ancestry-related structure.

\textbf{Association-based SNP selection}. 
Following the general QC and population structure assessment, we implemented a significance-based SNP filtering strategy for feature selection. SNPs were ranked according to their association p-values with the phenotype of interest. For the subsequent predictive modelling analyses, we retained the 39 most significantly associated SNPs, defined as those with p-values $ \leq 5.0 \times 10^{-5}.$

\clearpage
\bibliographystyle{apalike}
\bibliography{bib_file.bib}
\end{document}